\documentclass[journal]{IEEEtran}

\usepackage{array,booktabs}
\usepackage{algpseudocode}
\usepackage{algorithm}
\usepackage{amsfonts}
\usepackage{amsmath}
\usepackage{amssymb}
\usepackage{amsthm}
\usepackage{graphicx}
\usepackage[noadjust]{cite}
\usepackage{mathrsfs}
\usepackage{stmaryrd}
\usepackage{siunitx}
\usepackage{multicol}
\usepackage{tikz}
\usepackage{setspace}
\usetikzlibrary{shapes}

\usepackage{pst-plot,pst-node}
\usepackage{pstricks, pst-node, pst-plot, pst-circ}
\usepackage{moredefs}

\providelength{\AxesLineWidth}       
\providelength{\plotwidth}           
\providelength{\LineWidth}           
\providelength{\MarkerSize}          
\newrgbcolor{GridColor}{0.8 0.8 0.8}%

\newcommand{\revise}[1]{\textcolor{black}{#1}}

\DeclareGraphicsExtensions{.pdf}

\ifCLASSINFOpdf

\else

\fi

\begin{document}

\title{Design of {LDPC} Codes Robust to Noisy Message-Passing Decoding}

\author{Alla~Tarighati,~\IEEEmembership{ Student Member,~IEEE,}
        Hamed~Farhadi,~\IEEEmembership{Member,~IEEE}\\
        and~Farshad~Lahouti,~\IEEEmembership{Senior Member,~IEEE}

\thanks{A.~Tarighati is with the ACCESS Linnaeus Centre, department of signal processing,
KTH Royal Institute of Technology, Sweden (e-mail: allat@kth.se).}
\thanks{H.~Farhadi is with the Department of Signals and Systems, Chalmers University of Technology, Sweden (e-mail: farhadi@chalmers.se).}
\thanks{F.~Lahouti is with the Center for Wireless Multimedia Communications, School of Electrical and Computer Engineering, College of Engineering, University of Tehran, Iran (e-mail: lahouti@ut.ac.ir).}
\thanks{The material in this paper was presented in part at the 6th IEEE Int. Symp. Turbo Codes and Iterative Inform. Processing (ISTC), Brest, France, 2010.}
}

\maketitle

\begin{abstract}

We address noisy message-passing decoding of low-density parity-check (LDPC) codes over additive white Gaussian noise channels. \revise{Message-passing decoders in which certain processing units iteratively exchange messages are common for decoding LDPC codes. The exchanged messages are in general subject to internal noise in hardware implementation of these decoders.} We model the internal decoder noise as \revise{additive white Gaussian noise (AWGN)} degrading exchanged messages. Using Gaussian approximation of the exchanged messages, we perform a two-dimensional density evolution analysis for the noisy LDPC decoder. This \revise{makes it possible to track} both the mean, and the variance of the exchanged message densities, and hence, \revise{to quantify} the threshold of the LDPC code in the presence of internal decoder noise. The numerical and simulation results are presented that quantify the performance loss due to the internal decoder noise. To partially compensate this performance loss, we propose a simple method, based on EXIT chart analysis, to design robust irregular LDPC codes. The simulation results indicate that the designed codes can indeed compensate part of the performance loss due to the internal decoder noise.
\end{abstract}

%

\IEEEpeerreviewmaketitle

\section{Introduction}
\IEEEPARstart{L}{ow}-density parity-check (LDPC) codes -- first discovered by Gallager \cite{Gallag2} and rediscovered by Spielman \textit{et al.} \cite{spil} and MacKay \textit{et al.} \cite{mackay1}, \cite{mackay2} -- due to their outstanding performance have attracted much interest and have been studied  extensively during the recent years. They have also been included in several wireless communication standards, e.g., DVB-S2 and IEEE 802.16e \cite{dvb}, \cite{std}. Effective decoding of LDPC codes can be accomplished by using iterative message-passing schemes such as  belief-propagation algorithm or sum product algorithm \cite{chung01}. Moreover, \revise{powerful} analytical tools including \textit{EXIT chart analysis} \cite{ten04,ashikh04} and \textit{density evolution analysis} \cite{rich001} are developed for designing LDPC codes and quantifying their performance limits.

The sum-product algorithm is based on elementary computations using sum-product modules \cite{loeli01}.
These modules can be implemented using digital circuits or analog circuits.
The digital implementation of sum-product modules for LDPC decoding, e.g., the one presented in \cite{pego06}, is subject to noisy message passing due to the quantization of messages.
The impact of quantization error on LDPC decoding performance is investigated in \cite{zhang07}, where the simulations indicate the resulting substantial performance degradation. Quantization error is often modeled as an additive noise and is assumed Gaussian under certain conditions \cite{lee96}. In \cite{varsh11}, Varshney considers the impact of quantization error on message passing algorithm, and suggests a signal-independent additive truncated Gaussian noise model. More recently, Leduc-Primeau \emph{et al.} studied the impact of timing deviations (faults) on digital LDPC decoder circuits in \cite{leduc15}. They model the deviations as additive noise to the messages passed in the decoder, and show that under certain conditions the noise can be assumed Gaussian.

Loeliger \textit{et al.} in \cite{loeli01} and Hagenauer and Winklhofer in \cite{hagen2} introduced soft gates in order to implement sum-product modules using analog transistor circuits.
They have also shown that by \revise{the} variations of a single \revise{basic} circuit,
the entire family of sum-product modules can be implemented.
Using these circuits, any network of sum-product modules, in particular, iterative decoder of LDPC codes can be directly implemented in analog very-large-scale \revise{integrated} (VLSI) circuits.
In analog decoders, the exchanged messages are in general subject to additive intrinsic noise\revise{. The power of the noise} depends in part on the chip temperature \cite{koch09}.
To capture this phenomenon, in \cite{koch09} \revise{Koch \emph{et al.} considered a channel that is subject to an additive white Gaussian noise}.
This channel is motivated by point-to-point communication between two terminals that are embedded in the same chip.
Therefore, the \textit{internal decoder noise} may affect the communication of soft gates, and hence degrades the performance of the iterative analog decoder.

Since in practice digital or analog LDPC decoders are subject to internal noise, the impact of this noise on the performance of iterative decoding needs to be investigated.
Performance analysis of noisy LDPC decoding has attracted extensive interest recently (see e.g. \cite{varsh11,Hsi15,Taba12,Taba13,Huang14,us10} and the references therein).
The performance of a noisy bit-flipping LDPC decoder over a binary symmetric channel (BSC) is studied in \cite{varsh11}. In this setting, the decoder messages are exchanged over binary symmetric internal channels between the variable nodes and the check nodes. It has been shown that the performance degrades smoothly as the decoder noise probability increases.
Tabatabaei \textit{et al.} studied the performance limits of LDPC decoder when it is \revise{subject} to transient processor error \cite{Taba12,Taba13}.
This research was further generalized in \cite{Huang14} by considering both transient processor errors and permanent memory errors, using density evolution analysis for regular LDPC codes.

\begin{figure}
\begin{center}
\begin{tikzpicture}[align=center,scale=0.7,>=stealth] 
\node (v1) at (-2,2) [circle,minimum size=.4cm,draw] {}; \node at (-2.6,2) {$v_{1}$};
\node (v2) at (-2,0) [circle,minimum size=.4cm,draw] {}; \node at (-2.6,0) {$v_{2}$};
\node (vn) at (-2,-3) [circle,minimum size=.4cm,draw] {}; \node at (-2.6,-3) {$v_{N}$};
\node (c1) at (2,1.5) [rectangle,minimum size=.35cm,draw] {};  \node at (2.6,1.5) {$c_{1}$};
\node (c2) at (2,0) [rectangle,minimum size=.35cm,draw] {}; \node at (2.6,0) {$c_{2}$};
\node (ck) at (2,-2) [rectangle,minimum size=.35cm,draw] {}; \node at (2.6,-2) {$c_{K}$};
\draw [-] (v1) -- (-1,1.9) {};\draw [-] (v1) -- (-1.2,1.6) {};\draw [-] (v1) -- (-1.4,1.3) {};
\draw [-] (v2) -- (-1,0.3) {};\draw [-] (v2) -- (-1,0) {};\draw [-] (v2) -- (-1,-.3) {};
\draw [-] (vn) -- (-1,-2.9) {};\draw [-] (vn) -- (-1.2,-2.6) {};\draw [-] (vn) -- (-1.4,-2.3) {};
\draw [-] (c1) -- (1,1.6) {};\draw [-] (c1) -- (0.9,1.5) {};\draw [-] (c1) -- (0.8,1.4) {};
\draw [-] (c1) -- (1,1.1) {};\draw [-] (c1) -- (0.9,1.2) {};\draw [-] (c1) -- (0.8,1.3) {};
\draw [-] (c2) -- (1,0.3) {};\draw [-] (c2) -- (0.9,0.2) {};\draw [-] (c2) -- (0.8,.08) {};
\draw [-] (c2) -- (1,-0.3) {};\draw [-] (c2) -- (0.9,-0.2) {};\draw [-] (c2) -- (0.8,-.08) {};
\draw [-] (ck) -- (1,-2.1) {};\draw [-] (ck) -- (0.9,-2) {};\draw [-] (ck) -- (0.8,-1.9) {};
\draw [-] (ck) -- (1,-1.6) {};\draw [-] (ck) -- (0.9,-1.7) {};\draw [-] (ck) -- (0.8,-1.8) {};
\node[rotate=90] at (-2,-1.5) {$\dotsb$};\node[rotate=90] at (2,-1) {$\dotsb$};\node[rotate=90] at (0,0) {$\dotsb$};
\end{tikzpicture}
\end{center}
\caption{Tanner graph of a regular $(3,6)$ LDPC code, where squares denote check nodes and circles denote variable nodes.}
\label{graph}
\end{figure}
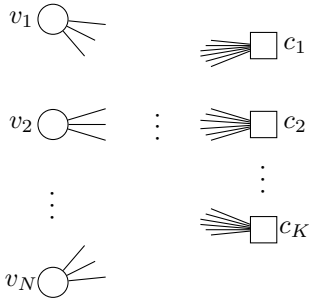

In this paper, we analyze the performance of LDPC codes \revise{transmitted} over an AWGN communication channel, when a sum-product decoding algorithm is employed in which exchanged messages are degraded by independent additive white Gaussian noise.
We first invoke a density evolution (DE) analysis to track the probability distribution of exchanged messages during decoding, and quantify the performance degradation due to the decoder noise.
We compute the density evolution equations for both regular and irregular LDPC codes.
Also, we introduce an algorithm to find the EXIT curves of a noisy decoder.
\revise{Finally we propose an} algorithm for the design of robust irregular LDPC codes using EXIT chart for noisy decoders to partially compensate the performance loss due to the internal decoder noise.

This paper is organized as follows.
In Section \ref{sec:principles}, we present the definitions and model for noisy message-passing decoder.
In Section \ref{sec:de}, \revise{we} derive the density evolution equations for the noisy message-passing decoder.
Next, numerical results of the density evolution analysis and simulation results of finite-length codes are presented.
In Section \ref{sec:exit}, EXIT chart analysis of the noisy decoder is presented.
 Using the EXIT charts, a method for designing robust codes for the noisy decoder is presented in Section \ref{sec:design}.
 Finally, Section \ref{sec:conclude} concludes the paper.

\section{LDPC Codes and Noisy Message-Passing Decoding Principles}\label{sec:principles}
Consider a regular binary $(d_{v},d_{c})$ LDPC code with length $N$\revise{. The code can be represented by a $K\times N$ parity check matrix $\mathbf{H}$ with binary elements, where the weight of each column and row of the matrix are $d_v$ and $d_c$, respectively. There is a Tanner graph corresponding to the parity check matrix} with $N$ variable nodes and $K\triangleq N\frac{d_{v}}{d_{c}}$ check nodes.
Every variable node in the \revise{graph} is connected to $d_v$ check nodes and every check node is connected to $d_c$ variable nodes.
Corresponding to the ones in the columns and \revise{the} rows of $\mathbf{H}$, the variable nodes and the check nodes are connected to each other in the Tanner graph. Fig. \ref{graph} exemplifies a Tanner graph for a regular $(3,6)$ LDPC code with length $N$.
Variable node $v_i$ and check node $c_j$ are known as \textit{neighbors}, if they are connected to each other.

Message-passing decoding algorithm can be represented as iterative exchange of messages between check nodes and variable nodes of the Tanner graph. Specifically, every check node receives messages from its $d_{c}$ neighbor variable nodes and sends the computed messages back.
Similarly, \revise{each variable node exchanges messages with its} $d_{v}$ neighbor check nodes.

We consider the output messages of the variable and check nodes as log-likelihood ratio (LLR) values,
where the sign of a variable node message specifies the bit estimate and its magnitude indicates the reliability of the estimation.

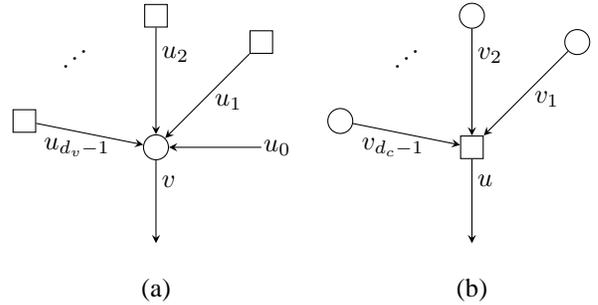
\begin{figure}[t]
\begin{center}
\begin{tikzpicture}[align=center,scale=0.7,>=stealth] 
\node (v) at (-3,-.5) [circle,minimum size=.3cm,draw] {};
\node (u1) at (-1,1.5) [rectangle,minimum size=.3cm,draw] {};
\node (u2) at (-3,2) [rectangle,minimum size=.3cm,draw] {};
\node (un) at (-5.5,0) [rectangle,minimum size=.3cm,draw] {};
\node (u) at (-3,-2.5) {}; 
\draw [->] (u1) -- (v) {} node [near start, below,inner sep=7pt] {$u_{1}$};
\draw [->] (u2) -- (v) {} node [near start, right,inner sep=2pt] {$u_{2}$};
\draw [->] (un) -- (v) {} ;\node at (-4.5,-0.5) {$u_{d_v-1}$};
\draw [->] (v) -- (u) {} node [near start, right,inner sep=2pt] {$v$};
\draw [->] (-1,-.5) -- (v); \node at (-.7, -0.5)  {$u_0$};
\node[rotate=40] at (-4.5,1.2) {$\dotsb$};\node at (-3,-3.2) {(a)};
\node (c) at (3,-.5) [rectangle,minimum size=.3cm,draw] {};
\node (v1) at (5,1.5) [circle,minimum size=.3cm,draw] {};
\node (v2) at (3,2) [circle,minimum size=.3cm,draw] {};
\node (vn) at (0.5,0) [circle,minimum size=.3cm,draw] {};
\node (v0) at (3,-2.5) {}; 
\draw [->] (v1) -- (c) {} node [near start, below,inner sep=7pt] {$v_{1}$};
\draw [->] (v2) -- (c) {}node [near start, right,inner sep=2pt] {$v_{2}$};
\draw [->] (vn) -- (c) {} ;\node at (1.5,-0.5) {$v_{d_c-1}$};
\draw [->] (c) -- (v0) {}node [near start, right,inner sep=2pt] {$u$};
\node[rotate=40] at (1.8,1.2) {$\dotsb$};\node at (3,-3.2) {(b)};
\end{tikzpicture}\vspace{-1.55mm}
\end{center}
\caption{Message flow through a variable node (a), and through a check node (b).}
\label{noiseless}
\end{figure}

According to the sum-product decoding algorithm, the message at iteration $l$ from a variable node to a check node, denoted by $v^{(l)}$, is
\begin{equation}
v^{(l)}=\sum_{i=0}^{d_{v}-1} u^{(l-1)}_{i},
\end{equation}
where $u_{i}^{(l-1)}$, $i=1,\ldots,d_{v}-1$, are incoming LLRs from variable node neighbors at iteration $l-1$, except the check node that is to receive the output message $v^{(l)}$, and $u_{0}$ is the incoming LLR message from the communication channel.
The message $u^{(l)}$ from a check node to a variable node at iteration $l$ can be obtained as follows
\begin{equation}
\tanh\frac{u^{(l)}}{2}=\prod^{d_{c}-1}_{j=1}\tanh\frac{v^{(l)}_{j}}{2}.
\end{equation}
Fig.~\ref{noiseless} shows the schematics of message-passing for a variable node and a check node.

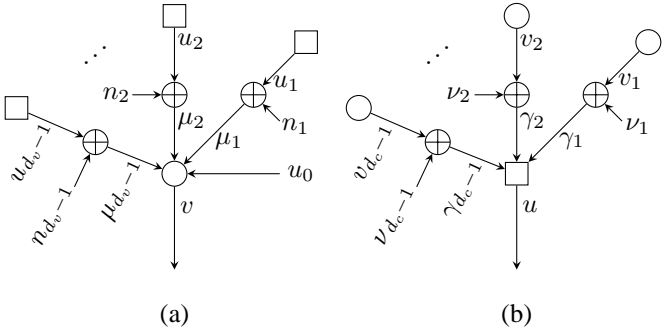
\begin{figure}[t]
\begin{center}
\begin{tikzpicture}[align=center,scale=0.7,>=stealth] 
\tikzset{oplus/.style={path picture={%
      \draw[black]
       (path picture bounding box.south) -- (path picture bounding box.north)
       (path picture bounding box.west) -- (path picture bounding box.east);
      }}}

\node (v) at (-4,-.5) [circle,minimum size=.3cm,draw] {};
\node (u1) at (-1.5,2) [rectangle,minimum size=.3cm,draw] {};
\node (u2) at (-4,2.5) [rectangle,minimum size=.3cm,draw] {};
\node (un) at (-7,0.75) [rectangle,minimum size=.3cm,draw] {};
\node (u) at (-4,-2.5) {}; 
\node (au1) at (-2.5,1) [oplus,draw,circle] {};
\node (au2) at (-4,1) [oplus,draw,circle] {};
\node (aun) at (-5.5,0.1) [oplus,draw,circle] {};
\node[rotate=40] at (-5.5,1.75) {$\ldots$};\node at (-4,-3.2) {(a)};
\draw [->] (u1) -- (au1) {} node [near start, below,inner sep=5pt] {$u_{1}$};
\draw [->] (au1) -- (v) {} node [near start, below,inner sep=5pt] {$\mu_{1}$};
\draw [->] (u2) -- (au2) {} node [near start, right,inner sep=1pt] {$u_{2}$};
\draw [->] (au2) -- (v) {} node [near start, right,inner sep=1pt] {$\mu_{2}$};
\draw [->] (un) -- (aun) {} ;\node[rotate=65] at (-5,-.8) {$\mu_{d_v-1}$};
\draw [->] (aun) -- (v) {};\node[rotate=65] at (-6.7,-0.1) {$u_{d_v-1}$};
\draw [->] (v) -- (u) {} node [near start, right,inner sep=2pt] {$v$};
\draw [->] (-2,-.5) -- (v); \node at (-1.6, -0.5)  {$u_0$};
\draw [->] (-2,0.5) -- (au1){};\node at (-1.7,0.4) {${n_{1}}$};
\draw [->] (-4.8,1) -- (au2){};\node at (-5.1,1) {$n_{{2}}$};
\draw [->] (-5.9,-.75) -- (aun){};\node [rotate=65] at (-6.3,-1.3) {$n_{{d_v-1}}$};

\node (c) at (2.5,-.5) [rectangle,minimum size=.3cm,draw] {};
\node (v1) at (5,2) [circle,minimum size=.3cm,draw] {};
\node (av1) at (4,1) [oplus,draw,circle] {};
\node (v2) at (2.5,2.5) [circle,minimum size=.3cm,draw] {};
\node (av2) at (2.5,1) [oplus,draw,circle] {};
\node (vn) at (-0.5,0.75) [circle,minimum size=.3cm,draw] {};
\node (avn) at (1,0.1) [oplus,draw,circle] {};
\node (v0) at (2.5,-2.5){}; 
\draw [->] (v1) -- (av1) {} node [near start, below,inner sep=5pt] {$v_{1}$};
\draw [->] (av1) -- (c) {} node [near start, below,inner sep=5pt] {$\gamma_{1}$};
\draw [->] (v2) -- (av2) {}node [near start, right,inner sep=1pt] {$v_{2}$};
\draw [->] (av2) -- (c) {}node [near start, right,inner sep=1pt] {$\gamma_{2}$};
\draw [->] (vn) -- (avn) {} ;\node[rotate=65] at (1.5,-.8) {$\gamma_{d_c-1}$};
\draw [->] (avn) -- (c) {};\node[rotate=65] at (-0.2,-0.1) {$v_{d_c-1}$};
\draw [->] (c) -- (v0) {}node [near start, right,inner sep=2pt] {$u$};
\node[rotate=40] at (1,1.75) {$\ldots$};\node at (2.5,-3.2) {(b)};
\draw [->] (4.5,.5) -- (av1){};\node at (4.8,0.4) {$\nu_{{1}}$};
\draw [->] (1.7,1) -- (av2){};\node at (1.4,1) {$\nu_{{2}}$};
\draw [->] (0.6,-.75) -- (avn){};\node [rotate=65] at (.2,-1.3) {$\nu{_{d_c-1}}$};
\end{tikzpicture}
\end{center}
\caption{The model of (a) a variable node (b) a check node in noisy belief-propagation decoder.}
\label{noisy}
\end{figure}

For a noisy decoder, the output messages of the variable and check nodes are subject to additive white Gaussian noise.
The conventional model shown in Fig.~\ref{noiseless} can be extended to the one in Fig.~\ref{noisy}, where \revise{${n_{i}}$ and ${\nu_{j}}$} denote the additive white Gaussian noise affecting the output messages of check nodes and variable nodes, respectively. Hence, ${\gamma_{j}}$ and ${\mu_{i}}$ are noisy versions of $v_{j}$ and $u_{i}$, respectively. Therefore, the incoming messages to the variable nodes and the check nodes are
\begin{equation}
\mu^{(l)}_{i}=u^{(l)}_{i}+\revise{n_{i}}\,,
\label{LLRincomingV}
\end{equation}
\begin{equation}
\gamma^{(l)}_{j}=v^{(l)}_{j}+\revise{\nu_{j}}\,,
\label{eq:gamma}\end{equation}
where \revise{$n_{i}$ and $\nu_{j}$} are assumed to be independent and identically distributed (i.i.d.), i.e.,  \revise{$n_{i},\nu_{j}\sim\mathcal{N}(0,\sigma_{d}^{2})$}.

According to the sum-product algorithm, at iteration $l$ the decoding is performed based on the following updating equations
\begin{equation}
v^{(l)}=u_{0}+\sum^{d_{v}-1}_{i=1}\mu^{(l-1)}_{i}\,,
\label{decoding Rule1}
\end{equation}
\begin{equation}
\tanh\frac{u^{(l)}}{2}=\prod^{d_{c}-1}_{j=1}\tanh\frac{\gamma^{(l)}_{j}}{2}\,.
\label{decoding Rule2}
\end{equation}

In order to generalize these equations to irregular case one can follow the same steps as the one described in \cite{chung01}.
In the next section, we propose an approach for the performance analysis of this noisy message-passing decoding algorithm.
\section{Density Evolution Analysis of Noisy Decoder}\label{sec:de}
In this \revise{section}, using Gaussian approximation, we will find the density evolution equations for the noisy decoder with regular variable and check degrees.
We further generalize the results to the irregular case,
and finally use the derived density evolution equations to evaluate the performance of noisy decoders.

\subsection{Gaussian Approximation and Consistency}
The density evolution analysis is an analytical method for tracking the densities of messages in iterative decoders.
This can be used to predict the performance limits of an LDPC code measured by code's threshold \cite{moon}.
The code's threshold is the smallest (largest) communication channel SNR (noise variance) for which an arbitrarily small decoding bit-error probability can be achieved by sufficiently long codewords.
For an AWGN communication channel and an LDPC sum-product decoder, the densities of the exchanged messages between the check nodes and the variable nodes can be approximated as Gaussian \cite{chung01,gamal01}.
Hence, these densities may be characterized only with their mean and variance.
A Gaussian random variable whose variance is twice its mean is said to be \textit{consistent} \cite{varsh11}.
The consistency assumption simplifies density evolution as a one-dimensional recursive equation based on the mean (or the variance) of the messages.
In \cite{chung01}, this assumption is used for the DE analysis of a \textit{noiseless} LDPC decoder,
and subsequently, quantifying the threshold of the code.

The key assumption in \revise{the} density evolution analysis of noise-free decoders is that the code block length is sufficiently large, based on which it may be assumed that the Tanner graph of the LDPC code is \revise{cycle-free}.
Since the code is linear and the communication channel is symmetric, we consider the transmission of an all-one codeword using a \revise{binary phase shift keying (BPSK)} modulation.
Thus, the \revise{LLR values} received over an AWGN communication channel are Gaussian distributed.
The mean and the variance of the received \revise{LLR values} are respectively equal to $m_{0}=\frac{2}{\sigma_{n}^{2}}$ and $\sigma_{0}^{2}=\frac{4}{\sigma_{n}^{2}}$, where $\sigma_{n}^{2}$ is the channel noise variance \cite{chung01}.
We assume that the \revise{random} variables $u$, $v$, $u_{i}$ and $v_{j}$ are all Gaussian \revise{distributed}.
First, we check whether the messages of a noisy sum-product LDPC decoder are consistent.
To this end, we consider the expected values of both sides of \eqref{LLRincomingV} and \eqref{decoding Rule1} and obtain
\begin{equation}
m_{v}^{(l)}=m_{0}+(d_{v}-1)m_{u}^{(l-1)},
\label{MVfinal}
\end{equation}
where $m_{v}^{(l)}$  and $m_{u}^{(l-1)}$ denote the means of output messages of variable nodes and check nodes, respectively. The index $i$ is omitted since $u_{i}$, $i=1,\dots, d_v-1$, are i.i.d. Next, by computing the variances of both sides of \eqref{LLRincomingV} we have
\begin{equation}
\revise{{\sigma^{2}_{\mu_{i}}}^{\!\!(l)}={\sigma^{2}_{u_{i}}}^{\!\!(l)}+\sigma^{2}_{d}}\,.
\label{eq:Vmu}\end{equation}
Using \eqref{decoding Rule1}, we obtain the variance of the variable node output
\begin{equation}
{\sigma^{2}_{v}}^{(l)}=\sigma^{2}_{0}
+\mathrm{var}\bigg(\sum_{i=1}^{d_{v}-1}\mu_{i}^{(l-1)}\bigg)
+2\mathrm{cov}\bigg(u_{0},\sum_{i=1}^{d_{v}-1}\mu_{i}^{(l-1)}\bigg),
\label{Vvariance}
\end{equation}
where $\mathrm{var}(X)$ denotes the variance of random variable $X$, and $\mathrm{cov}(X,Y)$ is the covariance of random variables $X$ and $Y$.
Since $\mu_{i}$, $i=1,\ldots,d_v-1$, are i.i.d.~Gaussian random variables, we have
\begin{align}
\mathrm{var}\bigg(\sum_{i=1}^{d_{v}-1}\mu_{i}^{(l-1)}\bigg)=
\sum_{i=1}^{d_{v}-1}\mathrm{var}\bigg(\mu_{i}^{(l-1)}\bigg)=
(d_{v}-1){\sigma^{2}_{\mu}}^{(l-1)}.
\end{align}

The last term in \eqref{Vvariance} is zero, as the Tanner graph of the code is assumed to be cycle-free and $u_{0}$ is independent of the noisy messages. Therefore, the variance of a variable node output message at iteration $l$ can be simplified as follows
\begin{equation}
{\sigma^{2}_{v}}^{(l)}=\sigma^{2}_{0}+(d_{v}-1){\sigma^{2}_{u}}^{(l-1)}+(d_{v}-1)\sigma^{2}_{d}\,.
\label{VVfinal}
\end{equation}

To verify the consistency of the noisy decoder, we plug in ${\sigma^{2}_{v}}^{(l)}=2m_v^{(l)}$ and ${\sigma^{2}_{u}}^{(l-1)}=2m_u^{(l-1)}$ into \eqref{VVfinal} and compare it with \eqref{MVfinal}.
It is clear that as long as $\sigma^{2}_{d}$ is non-zero, the \revise{two quantities} are not equal and hence the noisy LDPC decoder is \textit{not consistent}.
Therefore, it does not suffice to track only the mean values of the nodes' output messages.
Instead, it is required to track both the mean and the variance of nodes' output messages.
A similar situation \revise{has been} shown in the simulation results of \cite{saeedi07}, when there is an incorrect \revise{estimation} of the communication channel SNR at a (noiseless) LDPC decoder. However, since the code is linear and the communication channel is symmetric, sending all-one codeword is sufficient for statistical performance evaluation of the code \cite{saeedi07}.

For an irregular LDPC code, the degree distribution of variable nodes is $\lambda(x)=\sum_{i=2}^{D_{v}}\lambda_{i}x^{i-1}$ and that of the check nodes is  $\rho(x)=\sum_{i=2}^{D_{c}}\rho_{i}x^{i-1}$, where $\lambda_{i}$ and $\rho_{i}$ are the percentage of edges \revise{that are} connected to variable nodes and check nodes of degree $i$, respectively.
$D_{v}$ is the maximum degree of variable nodes and $D_c$ is the maximum degree of check nodes.
In this case, by similar steps as the ones for the regular LDPC codes, it can be shown that for the mean and the variance of a variable node of degree $i$ at iteration $l$
\begin{equation}
m_{v,i}^{(l)}=m_{0}+(i-1)m^{(l-1)}_{u},
\label{eq:irM}
\end{equation}
\begin{equation}
\revise{{\sigma^{2}_{v,i}}^{\!\!(l)}}=\sigma^{2}_{0}+(i-1){\sigma^{2}_{u}}^{(l-1)}+(i-1)\sigma^{2}_{d},
\label{eq:irV}
\end{equation}
where $m_{v,i}^{(l)}$ and ${\sigma^{2}_{v,i}}^{(l)}$ are the mean and the variance of a variable node of degree $i$ at iteration $l$, respectively.
From \eqref{eq:irM} and \eqref{eq:irV} it can be inferred that the consistency is not valid for the irregular case.
\begin{table}
\caption{Relation Between the Threshold and Decoder Noise Variance}
\begin{center}
\begin{tabular}[t]{|c|c|c|}
\hline
$\sigma^{2}_{d}$ & SNR threshold $\text{SNR}_\mathrm{th}$  & $(\sigma_{n})_\mathrm{th}$\\
\hline
\hline
$0$&$1.163$ dB&$0.8744$\\
\hline
$1$&$2.835$ dB&$0.7215$\\
\hline
$2$&$3.635$ dB&$0.6580$\\
\hline
$3$&$4.185$ dB&$0.6177$\\
\hline
\end{tabular}
\end{center}
\label{table1}
\end{table}
\subsection{Density Evolution with Gaussian Approximation for Noisy Message-Passing Decoder}
In the case of a noisy LDPC decoder, we have shown that consistency does not hold and we should track both the mean and the variance of nodes' output messages.
\revise{In order to do this}, we use the key equations \eqref{LLRincomingV}-\eqref{eq:Vmu} and \eqref{VVfinal}.
By computing the expected value of both sides of equation \eqref{decoding Rule2}, and noting that ${\gamma_j^{(l)},\, j=1,\ldots,d_c-1}$ are i.i.d., we have
\begin{align}
\begin{split}
\mathbb{E}\bigg[\tanh\frac{u^{(l)}}{2}\bigg]&=\mathbb{E}\bigg[ \prod^{d_{c}-1}_{j=1}\tanh\frac{\gamma^{(l)}_{j}}{2}\bigg]\\
&=\bigg(\mathbb{E}\bigg[\tanh\frac{\gamma^{(l)}}{2}\bigg]\bigg)^{d_{c}-1},
\label{eq:de1}\end{split}
\end{align}
where $\gamma^{(l)}$ \revise{is} defined in \eqref{eq:gamma} \revise{and} has the following distribution
\begin{equation}
\gamma^{(l)}\sim \mathcal{N}(m_{v}^{(l)},\,{\sigma^{2}_{v}}^{(l)}+\sigma_{d}^{2}).
\label{eq:gammadis}\end{equation}

Next, by computing the expected value of squared $\tanh$ rule, we obtain the second major equation as follows
\begin{equation}
\mathbb{E}\bigg[\tanh^{2}\left(\frac{u^{(l)}}{2}\right)\bigg]=\bigg(\mathbb{E}\bigg[\tanh^{2}\left(\frac{\gamma^{(l)}}{2}\right)\bigg]\bigg)^{d_{c}-1}.
\label{eq:de2}\end{equation}

The density evolution can be obtained by simultaneously solving equations \eqref{eq:de1} and \eqref{eq:de2}. Specifically, representing $\gamma^{(l)}$ using \eqref{eq:gammadis} and $m_{v}^{(l)}$, ${\sigma_{v}^{2}}^{(l)}$ from \eqref{MVfinal} and \eqref{VVfinal}, we obtain the DE equations for check nodes as follows
\begin{equation}\begin{split}
&f\left(m_{u}^{(l)},{\sigma^{2}_{u}}^{(l)}\right) =\left( f\left(m_{v}^{(l)},\,{\sigma_{v}^{2}}^{(l)}+\sigma^2_d\right)\right)^{d_c-1},\\
&g\left(m_{u}^{(l)},{\sigma^{2}_{u}}^{(l)}\right)=\left( g\left(m_{v}^{(l)},\,{\sigma_{v}^{2}}^{(l)}+\sigma^2_d\right)\right)^{d_c-1}.
\label{eq:DE}\end{split}\end{equation}
The auxiliary functions $f(m,\sigma^2)$ and $g(m,\sigma^2)$ are defined as follows
\begin{equation}\begin{split}
&f(m,\sigma^2)\triangleq \mathbb{E}\left[ \tanh\left(\frac{X}{2}\right)\right],\\
&g(m,\sigma^2)\triangleq \mathbb{E}\left[ \tanh^2 \left(\frac{X}{2}\right)\right],
\label{eq:aux}
\end{split}\end{equation}
where $X\sim\mathcal{N}(m,\sigma^{2})$.
These equations can be used to track $m_{u}^{(l)}$ and \revise{${\sigma_{u}^{2}}^{(l)}$} in the decoding iterations of a regular ($d_{v},d_{c}$) LDPC for given values of communication channel noise variance and internal decoder noise variance. \revise{Because of non-linearity of the equations in \eqref{eq:DE}, it is not easy to find a closed form expression for the mean and the variance at each iteration. As such, we resort to Monte-Carlo simulations to solve \eqref{eq:DE} and adopt the semi-Gaussian approximation method used in \cite{ardak04} and \cite{Hsi15}.}

Similarly, for irregular LDPC codes, the message distributions are approximated by Gaussian mixture \cite{chung01}\revise{. For} each check node of degree $i$ at iteration $l$ we have
\begin{equation}\begin{split}
&f\left(m_{u,i}^{(l)},{\sigma^{2}_{u,i}}^{(l)}\right) = \left[ \sum_{j=2}^{D_v}\lambda_j f\left(m_{v,j}^{(l)},\,\revise{{\sigma_{v,j}^{2}}^{\!\!\!(l)}}+\sigma^2_d\right)\right]^{i-1},\\
&g\left(m_{u,i}^{(l)},{\sigma^{2}_{u,i}}^{(l)}\right) =  \left[ \sum_{j=2}^{D_v}\lambda_j g\left(m_{v,j}^{(l)},\,\revise{{\sigma_{v,j}^{2}}^{\!\!\!(l)}}+\sigma^2_d\right)\right]^{i-1},\\
\label{eq:deirr}
\end{split}\end{equation}
and from Gaussian mixture equations, the density of check node in iteration $l$ has the following mean and variance values
\begin{equation}
m_{u}^{(l)}=\sum_{i=2}^{D_{c}}\rho_{i}m_{u,i}^{(l)}\,,
\label{Gmixmean}
\end{equation}
\begin{equation}
{\sigma_{u}^{2}}^{(l)}=\sum_{i=2}^{D_{c}}\rho_{i}\left[\revise{{\sigma_{u,i}^{2}}^{\!\!\!(l)}}+\left(m_{u,i}^{(l)}\right)^{2}\right] -\left(m_{u}^{(l)}\right)^{2}\,.
\label{Gmixvar}
\end{equation}
Therefore, the DE can be found by solving \eqref{eq:deirr} for a check node with degree $i$ and the distribution of a check node \revise{is then found} using \eqref{Gmixmean} and \eqref{Gmixvar}, iteratively.
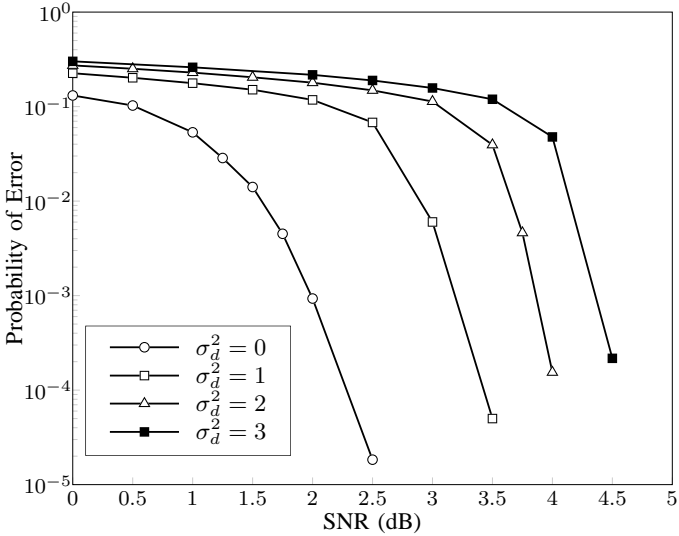
\begin{figure}[t]

%

%
\psset{xunit=0.200000\plotwidth,yunit=0.157742\plotwidth}%
\begin{pspicture}(-0.564516,-5.555556)(5.057604,0.116959)%


\psline[linewidth=\AxesLineWidth,linecolor=GridColor](0.000000,-5.000000)(0.000000,-4.923926)
\psline[linewidth=\AxesLineWidth,linecolor=GridColor](0.500000,-5.000000)(0.500000,-4.923926)
\psline[linewidth=\AxesLineWidth,linecolor=GridColor](1.000000,-5.000000)(1.000000,-4.923926)
\psline[linewidth=\AxesLineWidth,linecolor=GridColor](1.500000,-5.000000)(1.500000,-4.923926)
\psline[linewidth=\AxesLineWidth,linecolor=GridColor](2.000000,-5.000000)(2.000000,-4.923926)
\psline[linewidth=\AxesLineWidth,linecolor=GridColor](2.500000,-5.000000)(2.500000,-4.923926)
\psline[linewidth=\AxesLineWidth,linecolor=GridColor](3.000000,-5.000000)(3.000000,-4.923926)
\psline[linewidth=\AxesLineWidth,linecolor=GridColor](3.500000,-5.000000)(3.500000,-4.923926)
\psline[linewidth=\AxesLineWidth,linecolor=GridColor](4.000000,-5.000000)(4.000000,-4.923926)
\psline[linewidth=\AxesLineWidth,linecolor=GridColor](4.500000,-5.000000)(4.500000,-4.923926)
\psline[linewidth=\AxesLineWidth,linecolor=GridColor](5.000000,-5.000000)(5.000000,-4.923926)
\psline[linewidth=\AxesLineWidth,linecolor=GridColor](0.000000,-5.000000)(0.060000,-5.000000)
\psline[linewidth=\AxesLineWidth,linecolor=GridColor](0.000000,-4.000000)(0.060000,-4.000000)
\psline[linewidth=\AxesLineWidth,linecolor=GridColor](0.000000,-3.000000)(0.060000,-3.000000)
\psline[linewidth=\AxesLineWidth,linecolor=GridColor](0.000000,-2.000000)(0.060000,-2.000000)
\psline[linewidth=\AxesLineWidth,linecolor=GridColor](0.000000,-1.000000)(0.060000,-1.000000)
\psline[linewidth=\AxesLineWidth,linecolor=GridColor](0.000000,0.000000)(0.060000,0.000000)

\psline[linewidth=\AxesLineWidth,linecolor=GridColor](0.000000,-4.698970)(0.040000,-4.698970)
\psline[linewidth=\AxesLineWidth,linecolor=GridColor](0.000000,-4.522879)(0.040000,-4.522879)
\psline[linewidth=\AxesLineWidth,linecolor=GridColor](0.000000,-4.397940)(0.040000,-4.397940)
\psline[linewidth=\AxesLineWidth,linecolor=GridColor](0.000000,-4.301030)(0.040000,-4.301030)
\psline[linewidth=\AxesLineWidth,linecolor=GridColor](0.000000,-4.221849)(0.040000,-4.221849)
\psline[linewidth=\AxesLineWidth,linecolor=GridColor](0.000000,-4.154902)(0.040000,-4.154902)
\psline[linewidth=\AxesLineWidth,linecolor=GridColor](0.000000,-4.096910)(0.040000,-4.096910)
\psline[linewidth=\AxesLineWidth,linecolor=GridColor](0.000000,-4.045757)(0.040000,-4.045757)
\psline[linewidth=\AxesLineWidth,linecolor=GridColor](0.000000,-3.698970)(0.040000,-3.698970)
\psline[linewidth=\AxesLineWidth,linecolor=GridColor](0.000000,-3.522879)(0.040000,-3.522879)
\psline[linewidth=\AxesLineWidth,linecolor=GridColor](0.000000,-3.397940)(0.040000,-3.397940)
\psline[linewidth=\AxesLineWidth,linecolor=GridColor](0.000000,-3.301030)(0.040000,-3.301030)
\psline[linewidth=\AxesLineWidth,linecolor=GridColor](0.000000,-3.221849)(0.040000,-3.221849)
\psline[linewidth=\AxesLineWidth,linecolor=GridColor](0.000000,-3.154902)(0.040000,-3.154902)
\psline[linewidth=\AxesLineWidth,linecolor=GridColor](0.000000,-3.096910)(0.040000,-3.096910)
\psline[linewidth=\AxesLineWidth,linecolor=GridColor](0.000000,-3.045757)(0.040000,-3.045757)
\psline[linewidth=\AxesLineWidth,linecolor=GridColor](0.000000,-2.698970)(0.040000,-2.698970)
\psline[linewidth=\AxesLineWidth,linecolor=GridColor](0.000000,-2.522879)(0.040000,-2.522879)
\psline[linewidth=\AxesLineWidth,linecolor=GridColor](0.000000,-2.397940)(0.040000,-2.397940)
\psline[linewidth=\AxesLineWidth,linecolor=GridColor](0.000000,-2.301030)(0.040000,-2.301030)
\psline[linewidth=\AxesLineWidth,linecolor=GridColor](0.000000,-2.221849)(0.040000,-2.221849)
\psline[linewidth=\AxesLineWidth,linecolor=GridColor](0.000000,-2.154902)(0.040000,-2.154902)
\psline[linewidth=\AxesLineWidth,linecolor=GridColor](0.000000,-2.096910)(0.040000,-2.096910)
\psline[linewidth=\AxesLineWidth,linecolor=GridColor](0.000000,-2.045757)(0.040000,-2.045757)
\psline[linewidth=\AxesLineWidth,linecolor=GridColor](0.000000,-1.698970)(0.040000,-1.698970)
\psline[linewidth=\AxesLineWidth,linecolor=GridColor](0.000000,-1.522879)(0.040000,-1.522879)
\psline[linewidth=\AxesLineWidth,linecolor=GridColor](0.000000,-1.397940)(0.040000,-1.397940)
\psline[linewidth=\AxesLineWidth,linecolor=GridColor](0.000000,-1.301030)(0.040000,-1.301030)
\psline[linewidth=\AxesLineWidth,linecolor=GridColor](0.000000,-1.221849)(0.040000,-1.221849)
\psline[linewidth=\AxesLineWidth,linecolor=GridColor](0.000000,-1.154902)(0.040000,-1.154902)
\psline[linewidth=\AxesLineWidth,linecolor=GridColor](0.000000,-1.096910)(0.040000,-1.096910)
\psline[linewidth=\AxesLineWidth,linecolor=GridColor](0.000000,-1.045757)(0.040000,-1.045757)
\psline[linewidth=\AxesLineWidth,linecolor=GridColor](0.000000,-0.698970)(0.040000,-0.698970)
\psline[linewidth=\AxesLineWidth,linecolor=GridColor](0.000000,-0.522879)(0.040000,-0.522879)
\psline[linewidth=\AxesLineWidth,linecolor=GridColor](0.000000,-0.397940)(0.040000,-0.397940)
\psline[linewidth=\AxesLineWidth,linecolor=GridColor](0.000000,-0.301030)(0.040000,-0.301030)
\psline[linewidth=\AxesLineWidth,linecolor=GridColor](0.000000,-0.221849)(0.040000,-0.221849)
\psline[linewidth=\AxesLineWidth,linecolor=GridColor](0.000000,-0.154902)(0.040000,-0.154902)
\psline[linewidth=\AxesLineWidth,linecolor=GridColor](0.000000,-0.096910)(0.040000,-0.096910)
\psline[linewidth=\AxesLineWidth,linecolor=GridColor](0.000000,-0.045757)(0.040000,-0.045757)

{ \footnotesize 
\rput[t](0.000000,-5.076074){$0$}
\rput[t](0.500000,-5.076074){$0.5$}
\rput[t](1.000000,-5.076074){$1$}
\rput[t](1.500000,-5.076074){$1.5$}
\rput[t](2.000000,-5.076074){$2$}
\rput[t](2.500000,-5.076074){$2.5$}
\rput[t](3.000000,-5.076074){$3$}
\rput[t](3.500000,-5.076074){$3.5$}
\rput[t](4.000000,-5.076074){$4$}
\rput[t](4.500000,-5.076074){$4.5$}
\rput[t](5.000000,-5.076074){$5$}
\rput[r](0.00000,-5.000000){$10^{-5}$}
\rput[r](0.00000,-4.000000){$10^{-4}$}
\rput[r](0.00000,-3.000000){$10^{-3}$}
\rput[r](0.00000,-2.000000){$10^{-2}$}
\rput[r](0.00000,-1.000000){$10^{-1}$}
\rput[r](0.00000,0.000000){$10^{0}$}
} 

\psframe[linewidth=\AxesLineWidth,dimen=middle](0.000000,-5.000000)(5.000000,0.000000)

{ \small 
\rput[b](2.500000,-5.555556){
\begin{tabular}{c}
SNR (dB)\\
\end{tabular}
}

\rput[t]{90}(-0.574516,-2.500000){
\begin{tabular}{c}
Probability of Error\\
\end{tabular}
}
} 

\newrgbcolor{color587.001}{0  0  0}
\psline[plotstyle=line,linejoin=1,showpoints=true,dotstyle=Bo,dotsize=\MarkerSize,linestyle=solid,linewidth=\LineWidth,linecolor=color587.001]
(0.000000,-0.882229)(0.500000,-0.988644)(1.000000,-1.272306)(1.250000,-1.544180)(1.500000,-1.850874)
(1.750000,-2.345861)(2.000000,-3.030974)(2.500000,-4.735977)

\newrgbcolor{color588.0005}{0  0  0}
\psline[plotstyle=line,linejoin=1,showpoints=true,dotstyle=Bsquare,dotsize=\MarkerSize,linestyle=solid,linewidth=\LineWidth,linecolor=color588.0005]
(0.000000,-0.646899)(0.500000,-0.695024)(1.000000,-0.752163)(1.500000,-0.821331)(2.000000,-0.929010)
(2.500000,-1.167572)(3.000000,-2.220742)(3.500000,-4.301030)

\newrgbcolor{color589.0005}{0  0  0}
\psline[plotstyle=line,linejoin=1,showpoints=true,dotstyle=Btriangle,dotsize=\MarkerSize,linestyle=solid,linewidth=\LineWidth,linecolor=color589.0005]
(0.000000,-0.564159)(0.500000,-0.599620)(1.000000,-0.640112)(1.500000,-0.689889)(2.000000,-0.746839)
(2.500000,-0.828007)(3.000000,-0.946023)(3.500000,-1.404958)(3.750000,-2.335385)(4.000000,-3.811775)

\newrgbcolor{color590.0005}{0  0  0}
\psline[plotstyle=line,linejoin=1,showpoints=true,dotstyle=square*,dotsize=\MarkerSize,linestyle=solid,linewidth=\LineWidth,linecolor=color590.0005]
(0.000000,-0.521102)(1.000000,-0.584166)(2.000000,-0.663969)(2.500000,-0.723338)(3.000000,-0.803048)
(3.500000,-0.922401)(4.000000,-1.321007)(4.500000,-3.663702)

{ \small 
\rput(0.962993,-4.009668){%
\psshadowbox[framesep=0pt,shadowsize=0pt,linewidth=\AxesLineWidth]{\psframebox*{\begin{tabular}{l}
\Rnode{a1}{\hspace*{0.0ex}} \hspace*{0.7cm} \Rnode{a2}{~~$\sigma^2_d = 0$} \\
\Rnode{a3}{\hspace*{0.0ex}} \hspace*{0.7cm} \Rnode{a4}{~~$\sigma^2_d = 1$} \\
\Rnode{a5}{\hspace*{0.0ex}} \hspace*{0.7cm} \Rnode{a6}{~~$\sigma^2_d = 2$} \\
\Rnode{a7}{\hspace*{0.0ex}} \hspace*{0.7cm} \Rnode{a8}{~~$\sigma^2_d = 3$} \\
\end{tabular}}
\ncline[linestyle=solid,linewidth=\LineWidth,linecolor=color587.001]{a1}{a2} \ncput{\psdot[dotstyle=Bo,dotsize=\MarkerSize,linecolor=color587.001]}
\ncline[linestyle=solid,linewidth=\LineWidth,linecolor=color588.0005]{a3}{a4} \ncput{\psdot[dotstyle=Bsquare,dotsize=\MarkerSize,linecolor=color588.0005]}
\ncline[linestyle=solid,linewidth=\LineWidth,linecolor=color589.0005]{a5}{a6} \ncput{\psdot[dotstyle=Btriangle,dotsize=\MarkerSize,linecolor=color589.0005]}
\ncline[linestyle=solid,linewidth=\LineWidth,linecolor=color590.0005]{a7}{a8} \ncput{\psdot[dotstyle=square*,dotsize=\MarkerSize,linecolor=color590.0005]}
}%
}%
} 

\end{pspicture}%
\centering
\caption{Error probability performance of $(3,6)$ regular finite length code with decoder noise variance $\sigma_{d}^{2}$.}
\label{simtestDE}
\end{figure}
\subsection{Numerical Results of \revise{the Density Evolution Analysis}}
We solve the density evolution equations iteratively for a $(3,6)$ regular LDPC code considering $m_{u}^{(0)}=0$ and ${{\sigma_{u}^{2}}^{(0)}=0}$ as initial conditions.
This provides the mean and the variance of check nodes' outputs and allows for the computation of the threshold for the given variances of internal decoder noise and communication channel noise.

Table \ref{table1} shows the relation between the threshold and the internal decoder noise variance $\sigma^2_d$ resulting from \eqref{eq:DE}.
It can be observed that the SNR threshold $\text{SNR}_\mathrm{th}\triangleq (\frac{E_b}{N_0})_{\mathrm{th}}$ increases as the internal decoder noise variance increases.
This is in line with a similar observation in \cite{varsh11} on the performance of bit-flipping LDPC decoding in the presence of noisy message-passing over BSC channels, where the performance deteriorates as the cross-over probability of the internal decoder noise increases.

Simulation results confirm that our analytical results accurately predict the performance of finite-length codes as well.
Fig.~\ref{simtestDE} depicts the simulation results for the performance of a finite-length $(3,6)$ regular code with length $N=1008$.
It is evident that the threshold of this \revise{finite-length} code is fairly the same as our analytical threshold for various values of the internal decoder noise variance $\sigma^2_d$. One can use \eqref{Gmixmean} and \eqref{Gmixvar} to also track the density of irregular LDPC codes for noisy decoder.

In general, the presented analysis could be used to design irregular LDPC codes for noisy decoders.
Since the problem of designing \revise{irregular LDPC codes} by density evolution is not a convex problem, finding a good degree distribution requires complex computations and extensive search \cite{ardak04}.
Therefore, in the remaining sections after investigating the performance limits of the noisy decoder by means of EXIT chart, we will introduce a simple and effective method to design robust LDPC codes for the noisy decoder.
\section{EXIT Chart Analysis of Noisy Decoder}\label{sec:exit}
The EXIT chart analysis, first introduced in the pioneering work of ten Brink \cite{ten01}, is a powerful
tool for analyzing the performance of iterative turbo techniques.
It is mainly based on keeping track of the mutual information of channel input bits and variable node and check node outputs.

Let $X$ be a binary random variable denoting the BPSK modulated AWGN communication channel input which takes $\pm$1 values with equal probabilities.
If $f(y)$ is the probability density function (pdf) of the communication channel soft output $Y$, then, the mutual information of $X$ and $Y$ for a symmetric channel \cite{rich01,hagen} is
\begin{equation}
I(X;Y)=\frac{1}{2}\sum_{x=\pm1}\int_{-\infty}^{\infty}f(y|x)\log\left(\frac{f(y|x)}{f(y)}\right)dy\,.
\label{MI}
\end{equation}

In order to find the EXIT function of a noiseless decoder, the variable node and the check node EXIT functions \revise{can be computed using} a $J$-function \cite{ten04}, which \revise{directly results} from the consistency assumption of the decoder. Since this assumption is violated in the case of a noisy decoder, to compute the a priori and extrinsic mutual information, \revise{we compute these values according to the definition of mutual information.}
\begin{figure}[t]
\begin{minipage}[b]{0.45\linewidth}
\begin{tikzpicture}[align=center,scale=0.8,>=stealth]
\tikzset{oplus/.style={path picture={%
      \draw[black]
       (path picture bounding box.south) -- (path picture bounding box.north)
       (path picture bounding box.west) -- (path picture bounding box.east);
      }}}
\node (v) at (-0.5,0) [circle,minimum size=0.2cm,draw] {{\tiny{VND}}};
\node (pl1) at (-2,1) [oplus,draw,circle,thick]  {};
\node (pl2) at (-2,-1) [oplus,draw,circle,thick]  {};
\draw [->] (-3.75,1) -- (pl1) node [near start,above,inner sep=2pt] {$u_1$};
\draw [->] (-3.75,-1) -- (pl2) node [near start,below,inner sep=2pt] {$u_{d_v-1}$};
\draw [->] (pl1) -- (v) node [near start,below,inner sep=6pt] {};
\draw [->] (pl2) -- (v) node [near start,below,inner sep=6pt] {};
\draw [->]  (v) -- (1.25,0) node [near end,above,inner sep=2pt] {$v$};
\draw [->] (-0.5,2.7) -- (v) node [near start, right,inner sep=2pt] {$u_0$};
\draw [->] (-2,1.75) -- (pl1) node [near start,right,inner sep=1pt] {$n_1$};
\draw [->] (-2,-1.75) -- (pl2) node [near start,right,inner sep=1pt] {$n_{d_v-1}$};
\node at (-2,0) {$\vdots$};
\draw [line width=1, dashed] (0.7,-1.9) rectangle (-2.8,1.9);\node at (-.8,-2.2) {NVND};
\end{tikzpicture}
\end{minipage}
\hspace{0.5cm}
\begin{minipage}[b]{0.45\linewidth}
\begin{tikzpicture}[align=center,scale=0.8,>=stealth]
\tikzset{oplus/.style={path picture={%
      \draw[black]
       (path picture bounding box.south) -- (path picture bounding box.north)
       (path picture bounding box.west) -- (path picture bounding box.east);
      }}}
\node (v) at (-0.5,0) [rectangle,minimum size=0.6cm,draw] {{\tiny{CND}}};
\node (pl1) at (-2,1) [oplus,draw,circle,thick]  {};
\node (pl2) at (-2,-1) [oplus,draw,circle,thick]  {};
\draw [->] (-3.75,1) -- (pl1) node [near start,above,inner sep=2pt] {$v_1$};
\draw [->] (-3.75,-1) -- (pl2) node [near start,below,inner sep=2pt] {$v_{d_c-1}$};
\draw [->] (pl1) -- (v) node [near start,below,inner sep=6pt] {};
\draw [->] (pl2) -- (v) node [near start,below,inner sep=6pt] {};
\draw [->]  (v) -- (1.25,0) node [near end,above,inner sep=2pt] {$u$};
\draw [->] (-2,1.75) -- (pl1) node [near start,right,inner sep=1pt] {$\nu_1$};
\draw [->] (-2,-1.75) -- (pl2) node [near start,right,inner sep=1pt] {$\nu_{d_c-1}$};
\node at (-2,0) {$\vdots$};
\draw [line width=1, dashed] (0.7,-1.9) rectangle (-2.8,1.9);\node at (-.8,-2.2) {NCND};
\end{tikzpicture}
\end{minipage}
\caption{Modification of decoder components (VND or CND) to noisy decoder components (NVND or NCND).}
\label{modnode}
\end{figure}
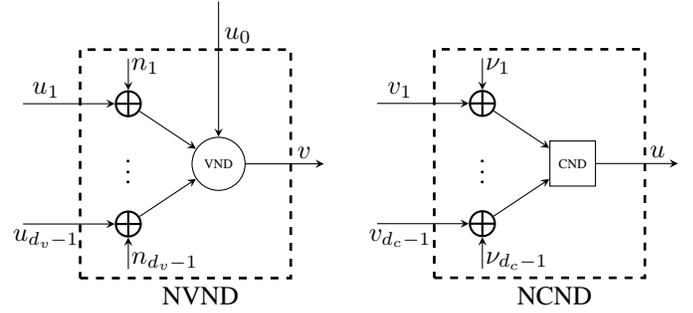
\begin{algorithm}[t]
\caption{EXIT curve for NVND}
\label{alg:exit}
\begin{algorithmic}[1]
\State \textbf{Input:} ${\textbf{I}}_\mathrm{A}$, $d_v$, $\sigma^2_n$, $\sigma^2_d$, $N$
\State \textbf{Output:} ${\textbf{I}}_\mathrm{E}$
\State {$\textbf{X}=\bf{1}$, [length $N$ BPSK modulated codeword]}
\For {$k=1:N$}
   [compute decoder LLR inputs]
   \begin{eqnarray}
   \textbf{Y}(k)=\frac{2}{\sigma^2_n}\left(1+{n}_c\right), n_c\sim \mathcal{N}(0,\sigma^2_n)\nonumber
   \end{eqnarray}
\EndFor
\For{$i = 1 : \text{length}({\textbf{I}}_\mathrm{A})$}
\begin{eqnarray}
\sigma_{A}=J^{-1}\left(\textbf{I}_{\mathrm{A}}(i)\right),\nonumber
\end{eqnarray}
where
\begin{eqnarray}
{J(\sigma)\triangleq1-\int_{-\infty}^{\infty}\frac{1}{\sqrt{2\pi}\sigma}e^{-\frac{(x-\sigma^2/2)^2}{2\sigma^2}}\,\log_2(1+e^{-x})\,dx}\nonumber
\end{eqnarray}
\For {$j=1:N$} [compute a priori LLRs]
\begin{eqnarray}
{\textbf{AP}(j)=\frac{\sigma_{A}^2}{2}+n, \, n\sim\mathcal{N}(0,\sigma_{A}^2)}\nonumber
\end{eqnarray}
\EndFor
\State [compute noisy variable node outputs using \eqref{decoding Rule1}]
\begin{eqnarray}
{\textbf{E}=\text{NVND}\left(d_v,\textbf{AP},\textbf{Y},\sigma^2_d\right)\nonumber }
\end{eqnarray}
\State [Compute extrinsic mutual information using \eqref{MI}]
\begin{eqnarray}
\textbf{I}_{\mathrm{E}}(i)={I}(\textbf{X},\textbf{E})\nonumber
\end{eqnarray}
\EndFor
\end{algorithmic}
\end{algorithm}

\begin{figure*}[t]
\centering
\begin{minipage}[b]{0.5\columnwidth}
\setlength{\plotwidth}{0.8\columnwidth}


%
\psset{xunit=1.000000\plotwidth,yunit=0.985887\plotwidth}%
\begin{pspicture}(-0.110599,0.111111)(1.011521,1.018713)%


\psline[linewidth=\AxesLineWidth,linecolor=GridColor](0.000000,0.200000)(0.000000,0.212172)
\psline[linewidth=\AxesLineWidth,linecolor=GridColor](0.100000,0.200000)(0.100000,0.212172)
\psline[linewidth=\AxesLineWidth,linecolor=GridColor](0.200000,0.200000)(0.200000,0.212172)
\psline[linewidth=\AxesLineWidth,linecolor=GridColor](0.300000,0.200000)(0.300000,0.212172)
\psline[linewidth=\AxesLineWidth,linecolor=GridColor](0.400000,0.200000)(0.400000,0.212172)
\psline[linewidth=\AxesLineWidth,linecolor=GridColor](0.500000,0.200000)(0.500000,0.212172)
\psline[linewidth=\AxesLineWidth,linecolor=GridColor](0.600000,0.200000)(0.600000,0.212172)
\psline[linewidth=\AxesLineWidth,linecolor=GridColor](0.700000,0.200000)(0.700000,0.212172)
\psline[linewidth=\AxesLineWidth,linecolor=GridColor](0.800000,0.200000)(0.800000,0.212172)
\psline[linewidth=\AxesLineWidth,linecolor=GridColor](0.900000,0.200000)(0.900000,0.212172)
\psline[linewidth=\AxesLineWidth,linecolor=GridColor](1.000000,0.200000)(1.000000,0.212172)
\psline[linewidth=\AxesLineWidth,linecolor=GridColor](0.000000,0.200000)(0.012000,0.200000)
\psline[linewidth=\AxesLineWidth,linecolor=GridColor](0.000000,0.300000)(0.012000,0.300000)
\psline[linewidth=\AxesLineWidth,linecolor=GridColor](0.000000,0.400000)(0.012000,0.400000)
\psline[linewidth=\AxesLineWidth,linecolor=GridColor](0.000000,0.500000)(0.012000,0.500000)
\psline[linewidth=\AxesLineWidth,linecolor=GridColor](0.000000,0.600000)(0.012000,0.600000)
\psline[linewidth=\AxesLineWidth,linecolor=GridColor](0.000000,0.700000)(0.012000,0.700000)
\psline[linewidth=\AxesLineWidth,linecolor=GridColor](0.000000,0.800000)(0.012000,0.800000)
\psline[linewidth=\AxesLineWidth,linecolor=GridColor](0.000000,0.900000)(0.012000,0.900000)
\psline[linewidth=\AxesLineWidth,linecolor=GridColor](0.000000,1.000000)(0.012000,1.000000)

{ \footnotesize 
\rput[t](0.000000,0.187828){$0$}
\rput[t](0.500000,0.187828){$0.5$}
\rput[t](1.000000,0.187828){$1$}
\rput[r](-0.012000,0.200000){$0.2$}
\rput[r](-0.012000,0.600000){$0.6$}
\rput[r](-0.012000,1.000000){$1$}
} 

\psframe[linewidth=\AxesLineWidth,dimen=middle](0.000000,0.200000)(1.000000,1.000000)

{ \small 
\rput[b](0.5,0.03){
\begin{tabular}{c}
$I_\mathrm{A,V}$ ,   $I_\mathrm{E,C}$\\
\end{tabular}
}

\rput[t]{90}(-0.23,0.600000){
\begin{tabular}{c}
$I_\mathrm{E,V}$ ,   $I_\mathrm{A,C}$\\
\end{tabular}
}
} 

\newrgbcolor{color38.012}{0  0  0}
\psline[plotstyle=line,linejoin=1,linestyle=solid,linewidth=\LineWidth,linecolor=color38.012]
(0.000698,0.200000)(0.000698,0.200000)
\psline[plotstyle=line,linejoin=1,linestyle=solid,linewidth=\LineWidth,linecolor=color38.012]
(0.000698,0.200000)(0.001471,0.240000)(0.002628,0.270000)(0.004966,0.310000)(0.005905,0.320000)
(0.006454,0.330000)(0.007707,0.340000)(0.008755,0.350000)(0.010178,0.360000)(0.011077,0.370000)
(0.013953,0.390000)(0.015631,0.400000)(0.017647,0.410000)(0.019401,0.420000)(0.021443,0.430000)
(0.023757,0.440000)(0.026920,0.450000)(0.029318,0.460000)(0.032056,0.470000)(0.035430,0.480000)
(0.039058,0.490000)(0.042460,0.500000)(0.046131,0.510000)(0.060389,0.540000)(0.076687,0.570000)
(0.090265,0.590000)(0.097601,0.600000)(0.104654,0.610000)(0.113884,0.620000)(0.121470,0.630000)
(0.130234,0.640000)(0.139860,0.650000)(0.150524,0.660000)(0.160602,0.670000)(0.171746,0.680000)
(0.196319,0.700000)(0.208369,0.710000)(0.222393,0.720000)(0.251826,0.740000)(0.267277,0.750000)
(0.283704,0.760000)(0.318416,0.780000)(0.337576,0.790000)(0.356188,0.800000)(0.396519,0.820000)
(0.420931,0.830000)(0.442727,0.840000)(0.467216,0.850000)(0.492918,0.860000)(0.518010,0.870000)
(0.573821,0.890000)(0.604026,0.900000)(0.634968,0.910000)(0.701646,0.930000)(0.739188,0.940000)
(0.775751,0.950000)(0.815745,0.960000)(0.904413,0.980000)(0.953670,0.990000)(1.000000,1.000000)

\newrgbcolor{color39.0115}{0  0  0}
\psline[plotstyle=line,linejoin=1,linestyle=dashed,linewidth=\LineWidth,linecolor=color39.0115]
(0.000000,0.719863)(0.030000,0.733582)(0.040000,0.738591)(0.060000,0.747639)(0.070000,0.751517)
(0.080000,0.756913)(0.090000,0.760833)(0.100000,0.765108)(0.130000,0.778708)(0.150000,0.787275)
(0.160000,0.790619)(0.170000,0.795102)(0.180000,0.800328)(0.190000,0.803403)(0.200000,0.808272)
(0.210000,0.812165)(0.220000,0.816453)(0.240000,0.824284)(0.250000,0.828541)(0.260000,0.832004)
(0.280000,0.841008)(0.290000,0.844655)(0.300000,0.847915)(0.310000,0.852011)(0.330000,0.859228)
(0.350000,0.866822)(0.360000,0.870269)(0.390000,0.879782)(0.400000,0.884118)(0.410000,0.887009)
(0.420000,0.890134)(0.430000,0.894142)(0.440000,0.896830)(0.470000,0.906921)(0.480000,0.909374)
(0.490000,0.913017)(0.510000,0.919254)(0.520000,0.921711)(0.550000,0.930447)(0.560000,0.933844)
(0.570000,0.936776)(0.580000,0.938749)(0.590000,0.942170)(0.600000,0.944627)(0.610000,0.946573)
(0.620000,0.949347)(0.630000,0.952378)(0.640000,0.953849)(0.650000,0.956776)(0.670000,0.961400)
(0.680000,0.963427)(0.690000,0.966021)(0.700000,0.967932)(0.710000,0.969572)(0.730000,0.973545)
(0.750000,0.976611)(0.770000,0.980502)(0.790000,0.983229)(0.800000,0.984899)(0.810000,0.986077)
(0.820000,0.987583)(0.840000,0.989771)(0.860000,0.992232)(0.880000,0.994197)(0.900000,0.995880)
(0.930000,0.997899)(0.950000,0.998911)(0.980000,0.999839)(0.990000,0.999967)

{ \scriptsize  
\rput(0.5,0.35){%
\psshadowbox[framesep=0pt,shadowsize=0pt,linewidth=\AxesLineWidth]{\psframebox*{\begin{tabular}{l}
\Rnode{a1}{\hspace*{0.0ex}} \hspace*{0.7cm} \Rnode{a2}{~~Check node} \\
\Rnode{a3}{\hspace*{0.0ex}} \hspace*{0.7cm} \Rnode{a4}{~~Variable node} \\
\end{tabular}}
\ncline[linestyle=solid,linewidth=\LineWidth,linecolor=color38.012]{a1}{a2}
\ncline[linestyle=dashed,linewidth=\LineWidth,linecolor=color39.0115]{a3}{a4}
}%
}%
} 

{ \small 
\newrgbcolor{color458.0114}{0  0  0}
\uput{0pt}[0](0.4855,0.755418){%
\psframebox[framesep=1pt,fillstyle=solid,linestyle=none,linewidth=0.5pt]{\begin{tabular}{@{}c@{}}
$\sigma^2_d = 0$\\[-0.3ex]
\end{tabular}}}
} 

\end{pspicture}%
\end{minipage}
\begin{minipage}[b]{0.5\columnwidth}
\setlength{\plotwidth}{0.8\columnwidth}

%
\psset{xunit=1.000000\plotwidth,yunit=0.985887\plotwidth}%
\begin{pspicture}(-0.110599,0.111111)(1.011521,1.018713)%


\psline[linewidth=\AxesLineWidth,linecolor=GridColor](0.000000,0.200000)(0.000000,0.212172)
\psline[linewidth=\AxesLineWidth,linecolor=GridColor](0.100000,0.200000)(0.100000,0.212172)
\psline[linewidth=\AxesLineWidth,linecolor=GridColor](0.200000,0.200000)(0.200000,0.212172)
\psline[linewidth=\AxesLineWidth,linecolor=GridColor](0.300000,0.200000)(0.300000,0.212172)
\psline[linewidth=\AxesLineWidth,linecolor=GridColor](0.400000,0.200000)(0.400000,0.212172)
\psline[linewidth=\AxesLineWidth,linecolor=GridColor](0.500000,0.200000)(0.500000,0.212172)
\psline[linewidth=\AxesLineWidth,linecolor=GridColor](0.600000,0.200000)(0.600000,0.212172)
\psline[linewidth=\AxesLineWidth,linecolor=GridColor](0.700000,0.200000)(0.700000,0.212172)
\psline[linewidth=\AxesLineWidth,linecolor=GridColor](0.800000,0.200000)(0.800000,0.212172)
\psline[linewidth=\AxesLineWidth,linecolor=GridColor](0.900000,0.200000)(0.900000,0.212172)
\psline[linewidth=\AxesLineWidth,linecolor=GridColor](1.000000,0.200000)(1.000000,0.212172)
\psline[linewidth=\AxesLineWidth,linecolor=GridColor](0.000000,0.200000)(0.012000,0.200000)
\psline[linewidth=\AxesLineWidth,linecolor=GridColor](0.000000,0.300000)(0.012000,0.300000)
\psline[linewidth=\AxesLineWidth,linecolor=GridColor](0.000000,0.400000)(0.012000,0.400000)
\psline[linewidth=\AxesLineWidth,linecolor=GridColor](0.000000,0.500000)(0.012000,0.500000)
\psline[linewidth=\AxesLineWidth,linecolor=GridColor](0.000000,0.600000)(0.012000,0.600000)
\psline[linewidth=\AxesLineWidth,linecolor=GridColor](0.000000,0.700000)(0.012000,0.700000)
\psline[linewidth=\AxesLineWidth,linecolor=GridColor](0.000000,0.800000)(0.012000,0.800000)
\psline[linewidth=\AxesLineWidth,linecolor=GridColor](0.000000,0.900000)(0.012000,0.900000)
\psline[linewidth=\AxesLineWidth,linecolor=GridColor](0.000000,1.000000)(0.012000,1.000000)

{ \footnotesize 
\rput[t](0.000000,0.187828){$0$}
\rput[t](0.500000,0.187828){$0.5$}
\rput[t](1.000000,0.187828){$1$}
\rput[r](-0.012000,0.200000){$0.2$}
\rput[r](-0.012000,0.600000){$0.6$}
\rput[r](-0.012000,1.000000){$1$}
} 

\psframe[linewidth=\AxesLineWidth,dimen=middle](0.000000,0.200000)(1.000000,1.000000)

{ \small 
\rput[b](0.5,0.03){
\begin{tabular}{c}
$I_\mathrm{A,V}$ ,   $I_\mathrm{E,C}$\\
\end{tabular}
}

\rput[t]{90}(-0.23,0.600000){
\begin{tabular}{c}
$I_\mathrm{E,V}$ ,   $I_\mathrm{A,C}$\\
\end{tabular}
}
} 

\newrgbcolor{color47.0128}{0  0  0}
\psline[plotstyle=line,linejoin=1,linestyle=solid,linewidth=\LineWidth,linecolor=color47.0128]
(0.000064,0.200000)(0.000064,0.200000)
\psline[plotstyle=line,linejoin=1,linestyle=solid,linewidth=\LineWidth,linecolor=color47.0128]
(0.000064,0.200000)(0.000289,0.250000)(0.000947,0.300000)(0.002144,0.340000)(0.004193,0.380000)
(0.006830,0.410000)(0.007819,0.420000)(0.009140,0.430000)(0.010230,0.440000)(0.013667,0.460000)
(0.015158,0.470000)(0.019333,0.490000)(0.022162,0.500000)(0.024742,0.510000)(0.028093,0.520000)
(0.030974,0.530000)(0.034470,0.540000)(0.038202,0.550000)(0.042640,0.560000)(0.046806,0.570000)
(0.052007,0.580000)(0.063544,0.600000)(0.069118,0.610000)(0.075400,0.620000)(0.090772,0.640000)
(0.098869,0.650000)(0.107624,0.660000)(0.116925,0.670000)(0.127003,0.680000)(0.148385,0.700000)
(0.172311,0.720000)(0.187386,0.730000)(0.199334,0.740000)(0.247587,0.770000)(0.282727,0.790000)
(0.303583,0.800000)(0.323179,0.810000)(0.345616,0.820000)(0.366855,0.830000)(0.415705,0.850000)
(0.440709,0.860000)(0.469763,0.870000)(0.497686,0.880000)(0.559448,0.900000)(0.593011,0.910000)
(0.666870,0.930000)(0.705584,0.940000)(0.748538,0.950000)(0.793850,0.960000)(0.842252,0.970000)
(0.892575,0.980000)(0.946861,0.990000)(1.000000,1.000000)

\newrgbcolor{color48.0123}{0  0  0}
\psline[plotstyle=line,linejoin=1,linestyle=dashed,linewidth=\LineWidth,linecolor=color48.0123]
(0.000000,0.644678)(0.020000,0.655966)(0.030000,0.661288)(0.040000,0.667466)(0.050000,0.672168)
(0.060000,0.678655)(0.070000,0.683633)(0.090000,0.695301)(0.100000,0.701208)(0.120000,0.711280)
(0.130000,0.717703)(0.140000,0.722552)(0.150000,0.729026)(0.160000,0.733444)(0.170000,0.738770)
(0.190000,0.750016)(0.200000,0.755314)(0.220000,0.765052)(0.230000,0.770229)(0.240000,0.774720)
(0.250000,0.780289)(0.260000,0.784919)(0.270000,0.791288)(0.280000,0.795583)(0.290000,0.800911)
(0.300000,0.805354)(0.310000,0.810447)(0.320000,0.814707)(0.330000,0.820353)(0.340000,0.824116)
(0.350000,0.829204)(0.360000,0.833751)(0.380000,0.841790)(0.400000,0.851272)(0.420000,0.859112)
(0.430000,0.863801)(0.440000,0.867716)(0.450000,0.872529)(0.470000,0.879620)(0.480000,0.884310)
(0.500000,0.892281)(0.510000,0.896815)(0.520000,0.899557)(0.540000,0.907538)(0.550000,0.911291)
(0.560000,0.914825)(0.570000,0.917984)(0.580000,0.922266)(0.590000,0.924852)(0.600000,0.928264)
(0.610000,0.932310)(0.620000,0.935240)(0.630000,0.938478)(0.640000,0.941163)(0.660000,0.947610)
(0.670000,0.950373)(0.680000,0.952622)(0.690000,0.955750)(0.710000,0.961047)(0.720000,0.963065)
(0.730000,0.966208)(0.750000,0.970221)(0.770000,0.974603)(0.780000,0.976569)(0.790000,0.978312)
(0.800000,0.980755)(0.810000,0.982615)(0.820000,0.983702)(0.840000,0.987271)(0.860000,0.989987)
(0.900000,0.994482)(0.930000,0.997275)(0.960000,0.999024)(0.980000,0.999775)(0.990000,0.999964)

{ \scriptsize  
\rput(0.5,0.35){%
\psshadowbox[framesep=0pt,shadowsize=0pt,linewidth=\AxesLineWidth]{\psframebox*{\begin{tabular}{l}
\Rnode{a1}{\hspace*{0.0ex}} \hspace*{0.7cm} \Rnode{a2}{~~Check node} \\
\Rnode{a3}{\hspace*{0.0ex}} \hspace*{0.7cm} \Rnode{a4}{~~Variable node} \\
\end{tabular}}
\ncline[linestyle=solid,linewidth=\LineWidth,linecolor=color47.0128]{a1}{a2}
\ncline[linestyle=dashed,linewidth=\LineWidth,linecolor=color48.0123]{a3}{a4}
}%
}%
} 

{ \small 
\newrgbcolor{color458.0122}{0  0  0}
\uput{0pt}[0](0.4855,0.755418){%
\psframebox[framesep=1pt,fillstyle=solid,linestyle=none,linewidth=0.5pt]{\begin{tabular}{@{}c@{}}
$\sigma^2_d = 1$\\[-0.3ex]
\end{tabular}}}
} 

\end{pspicture}%
\end{minipage}
\begin{minipage}[b]{0.5\columnwidth}
\setlength{\plotwidth}{0.8\columnwidth}
\psset{xunit=1.000000\plotwidth,yunit=0.985887\plotwidth}%
\begin{pspicture}(-0.110599,0.111111)(1.011521,1.018713)%


\psline[linewidth=\AxesLineWidth,linecolor=GridColor](0.000000,0.200000)(0.000000,0.212172)
\psline[linewidth=\AxesLineWidth,linecolor=GridColor](0.100000,0.200000)(0.100000,0.212172)
\psline[linewidth=\AxesLineWidth,linecolor=GridColor](0.200000,0.200000)(0.200000,0.212172)
\psline[linewidth=\AxesLineWidth,linecolor=GridColor](0.300000,0.200000)(0.300000,0.212172)
\psline[linewidth=\AxesLineWidth,linecolor=GridColor](0.400000,0.200000)(0.400000,0.212172)
\psline[linewidth=\AxesLineWidth,linecolor=GridColor](0.500000,0.200000)(0.500000,0.212172)
\psline[linewidth=\AxesLineWidth,linecolor=GridColor](0.600000,0.200000)(0.600000,0.212172)
\psline[linewidth=\AxesLineWidth,linecolor=GridColor](0.700000,0.200000)(0.700000,0.212172)
\psline[linewidth=\AxesLineWidth,linecolor=GridColor](0.800000,0.200000)(0.800000,0.212172)
\psline[linewidth=\AxesLineWidth,linecolor=GridColor](0.900000,0.200000)(0.900000,0.212172)
\psline[linewidth=\AxesLineWidth,linecolor=GridColor](1.000000,0.200000)(1.000000,0.212172)
\psline[linewidth=\AxesLineWidth,linecolor=GridColor](0.000000,0.200000)(0.012000,0.200000)
\psline[linewidth=\AxesLineWidth,linecolor=GridColor](0.000000,0.300000)(0.012000,0.300000)
\psline[linewidth=\AxesLineWidth,linecolor=GridColor](0.000000,0.400000)(0.012000,0.400000)
\psline[linewidth=\AxesLineWidth,linecolor=GridColor](0.000000,0.500000)(0.012000,0.500000)
\psline[linewidth=\AxesLineWidth,linecolor=GridColor](0.000000,0.600000)(0.012000,0.600000)
\psline[linewidth=\AxesLineWidth,linecolor=GridColor](0.000000,0.700000)(0.012000,0.700000)
\psline[linewidth=\AxesLineWidth,linecolor=GridColor](0.000000,0.800000)(0.012000,0.800000)
\psline[linewidth=\AxesLineWidth,linecolor=GridColor](0.000000,0.900000)(0.012000,0.900000)
\psline[linewidth=\AxesLineWidth,linecolor=GridColor](0.000000,1.000000)(0.012000,1.000000)

{ \footnotesize 
\rput[t](0.000000,0.187828){$0$}
\rput[t](0.500000,0.187828){$0.5$}
\rput[t](1.000000,0.187828){$1$}
\rput[r](-0.012000,0.200000){$0.2$}
\rput[r](-0.012000,0.600000){$0.6$}
\rput[r](-0.012000,1.000000){$1$}
} 

\psframe[linewidth=\AxesLineWidth,dimen=middle](0.000000,0.200000)(1.000000,1.000000)

{ \small 
\rput[b](0.5,0.03){
\begin{tabular}{c}
$I_\mathrm{A,V}$ ,   $I_\mathrm{E,C}$\\
\end{tabular}
}

\rput[t]{90}(-0.23,0.600000){
\begin{tabular}{c}
$I_\mathrm{E,V}$ ,   $I_\mathrm{A,C}$\\
\end{tabular}
}
} 

\newrgbcolor{color268.014}{0  0  0}
\psline[plotstyle=line,linejoin=1,linestyle=solid,linewidth=\LineWidth,linecolor=color268.014]
(0.000015,0.200000)(0.000015,0.200000)
\psline[plotstyle=line,linejoin=1,linestyle=solid,linewidth=\LineWidth,linecolor=color268.014]
(0.000015,0.200000)(0.000324,0.300000)(0.000820,0.340000)(0.001825,0.380000)(0.002499,0.400000)
(0.004936,0.440000)(0.009129,0.480000)(0.012373,0.500000)(0.016111,0.520000)(0.018560,0.530000)
(0.020692,0.540000)(0.023721,0.550000)(0.027025,0.560000)(0.029802,0.570000)(0.037478,0.590000)
(0.046974,0.610000)(0.052913,0.620000)(0.058043,0.630000)(0.064327,0.640000)(0.070872,0.650000)
(0.078497,0.660000)(0.085779,0.670000)(0.094618,0.680000)(0.123882,0.710000)(0.135061,0.720000)
(0.160698,0.740000)(0.174230,0.750000)(0.189157,0.760000)(0.204562,0.770000)(0.239351,0.790000)
(0.257705,0.800000)(0.277635,0.810000)(0.298066,0.820000)(0.322038,0.830000)(0.343781,0.840000)
(0.370080,0.850000)(0.424503,0.870000)(0.454229,0.880000)(0.485143,0.890000)(0.555051,0.910000)
(0.632154,0.930000)(0.675568,0.940000)(0.720487,0.950000)(0.768678,0.960000)(0.878248,0.980000)
(0.941132,0.990000)(1.000000,1.000000)

\newrgbcolor{color269.0135}{0  0  0}
\psline[plotstyle=line,linejoin=1,linestyle=dashed,linewidth=\LineWidth,linecolor=color269.0135]
(0.000000,0.582000)(0.020000,0.595277)(0.030000,0.600930)(0.070000,0.626254)(0.080000,0.632389)
(0.100000,0.645151)(0.110000,0.651745)(0.150000,0.675051)(0.160000,0.682687)(0.170000,0.688710)
(0.180000,0.693964)(0.190000,0.700120)(0.200000,0.705918)(0.210000,0.713400)(0.230000,0.724141)
(0.240000,0.731015)(0.250000,0.735805)(0.260000,0.742974)(0.270000,0.747906)(0.280000,0.754007)
(0.290000,0.758503)(0.300000,0.765118)(0.310000,0.770717)(0.320000,0.775934)(0.330000,0.781581)
(0.340000,0.787895)(0.350000,0.793437)(0.360000,0.797960)(0.380000,0.808076)(0.390000,0.813871)
(0.400000,0.817886)(0.410000,0.822988)(0.420000,0.828510)(0.450000,0.843382)(0.460000,0.847900)
(0.470000,0.853810)(0.480000,0.858974)(0.500000,0.868443)(0.520000,0.876832)(0.530000,0.881921)
(0.540000,0.886113)(0.550000,0.890626)(0.560000,0.895506)(0.570000,0.899208)(0.580000,0.903794)
(0.600000,0.911774)(0.610000,0.916260)(0.620000,0.919371)(0.630000,0.923853)(0.640000,0.926813)
(0.650000,0.931458)(0.670000,0.937914)(0.680000,0.941421)(0.690000,0.945279)(0.700000,0.948132)
(0.710000,0.951840)(0.720000,0.953673)(0.730000,0.957532)(0.750000,0.963245)(0.760000,0.965392)
(0.780000,0.971237)(0.800000,0.975393)(0.810000,0.977807)(0.820000,0.979826)(0.830000,0.982059)
(0.840000,0.983732)(0.860000,0.987631)(0.870000,0.988818)(0.900000,0.993240)(0.930000,0.996584)
(0.950000,0.998159)(0.970000,0.999289)(0.990000,0.999928)

{ \scriptsize  
\rput(0.5,0.35){%
\psshadowbox[framesep=0pt,shadowsize=0pt,linewidth=\AxesLineWidth]{\psframebox*{\begin{tabular}{l}
\Rnode{a1}{\hspace*{0.0ex}} \hspace*{0.7cm} \Rnode{a2}{~~Check node} \\
\Rnode{a3}{\hspace*{0.0ex}} \hspace*{0.7cm} \Rnode{a4}{~~Variable node} \\
\end{tabular}}
\ncline[linestyle=solid,linewidth=\LineWidth,linecolor=color268.014]{a1}{a2}
\ncline[linestyle=dashed,linewidth=\LineWidth,linecolor=color269.0135]{a3}{a4}
}%
}%
} 

{ \small 
\newrgbcolor{color200.0131}{0  0  0}
\uput{0pt}[0](0.4855,0.755418){%
\psframebox[framesep=1pt,fillstyle=solid,linestyle=none,linewidth=0.5pt]{\begin{tabular}{@{}c@{}}
$\sigma^2_d = 2$\\[-0.3ex]
\end{tabular}}}
} 

\end{pspicture}%
\end{minipage}
\begin{minipage}[b]{0.5\columnwidth}
\setlength{\plotwidth}{0.8\columnwidth}%
\psset{xunit=1.000000\plotwidth,yunit=0.985887\plotwidth}%
\begin{pspicture}(-0.110599,0.111111)(1.011521,1.018713)%


\psline[linewidth=\AxesLineWidth,linecolor=GridColor](0.000000,0.200000)(0.000000,0.212172)
\psline[linewidth=\AxesLineWidth,linecolor=GridColor](0.100000,0.200000)(0.100000,0.212172)
\psline[linewidth=\AxesLineWidth,linecolor=GridColor](0.200000,0.200000)(0.200000,0.212172)
\psline[linewidth=\AxesLineWidth,linecolor=GridColor](0.300000,0.200000)(0.300000,0.212172)
\psline[linewidth=\AxesLineWidth,linecolor=GridColor](0.400000,0.200000)(0.400000,0.212172)
\psline[linewidth=\AxesLineWidth,linecolor=GridColor](0.500000,0.200000)(0.500000,0.212172)
\psline[linewidth=\AxesLineWidth,linecolor=GridColor](0.600000,0.200000)(0.600000,0.212172)
\psline[linewidth=\AxesLineWidth,linecolor=GridColor](0.700000,0.200000)(0.700000,0.212172)
\psline[linewidth=\AxesLineWidth,linecolor=GridColor](0.800000,0.200000)(0.800000,0.212172)
\psline[linewidth=\AxesLineWidth,linecolor=GridColor](0.900000,0.200000)(0.900000,0.212172)
\psline[linewidth=\AxesLineWidth,linecolor=GridColor](1.000000,0.200000)(1.000000,0.212172)
\psline[linewidth=\AxesLineWidth,linecolor=GridColor](0.000000,0.200000)(0.012000,0.200000)
\psline[linewidth=\AxesLineWidth,linecolor=GridColor](0.000000,0.300000)(0.012000,0.300000)
\psline[linewidth=\AxesLineWidth,linecolor=GridColor](0.000000,0.400000)(0.012000,0.400000)
\psline[linewidth=\AxesLineWidth,linecolor=GridColor](0.000000,0.500000)(0.012000,0.500000)
\psline[linewidth=\AxesLineWidth,linecolor=GridColor](0.000000,0.600000)(0.012000,0.600000)
\psline[linewidth=\AxesLineWidth,linecolor=GridColor](0.000000,0.700000)(0.012000,0.700000)
\psline[linewidth=\AxesLineWidth,linecolor=GridColor](0.000000,0.800000)(0.012000,0.800000)
\psline[linewidth=\AxesLineWidth,linecolor=GridColor](0.000000,0.900000)(0.012000,0.900000)
\psline[linewidth=\AxesLineWidth,linecolor=GridColor](0.000000,1.000000)(0.012000,1.000000)

{ \footnotesize 
\rput[t](0.000000,0.187828){$0$}
\rput[t](0.500000,0.187828){$0.5$}
\rput[t](1.000000,0.187828){$1$}
\rput[r](-0.012000,0.200000){$0.2$}
\rput[r](-0.012000,0.600000){$0.6$}
\rput[r](-0.012000,1.000000){$1$}
} 

\psframe[linewidth=\AxesLineWidth,dimen=middle](0.000000,0.200000)(1.000000,1.000000)

{ \small 
\rput[b](0.5,0.03){
\begin{tabular}{c}
$I_\mathrm{A,V}$ ,   $I_\mathrm{E,C}$\\
\end{tabular}
}

\rput[t]{90}(-0.23,0.600000){
\begin{tabular}{c}
$I_\mathrm{E,V}$ ,   $I_\mathrm{A,C}$\\
\end{tabular}
}
} 

\newrgbcolor{color268.015}{0  0  0}
\psline[plotstyle=line,linejoin=1,linestyle=solid,linewidth=\LineWidth,linecolor=color268.015]
(0.000010,0.200000)(0.000010,0.200000)
\psline[plotstyle=line,linejoin=1,linestyle=solid,linewidth=\LineWidth,linecolor=color268.015]
(0.000010,0.200000)(0.000169,0.310000)(0.000713,0.370000)(0.001303,0.400000)(0.001825,0.420000)
(0.003228,0.450000)(0.004540,0.470000)(0.006282,0.490000)(0.007447,0.500000)(0.008346,0.510000)
(0.009824,0.520000)(0.013223,0.540000)(0.015079,0.550000)(0.019869,0.570000)(0.022669,0.580000)
(0.028998,0.600000)(0.036828,0.620000)(0.041476,0.630000)(0.057267,0.660000)(0.064803,0.670000)
(0.071455,0.680000)(0.088081,0.700000)(0.096924,0.710000)(0.118003,0.730000)(0.130485,0.740000)
(0.142249,0.750000)(0.155310,0.760000)(0.169812,0.770000)(0.185595,0.780000)(0.202253,0.790000)
(0.219530,0.800000)(0.258416,0.820000)(0.281246,0.830000)(0.304605,0.840000)(0.356770,0.860000)
(0.384406,0.870000)(0.414153,0.880000)(0.446636,0.890000)(0.480551,0.900000)(0.517339,0.910000)
(0.557070,0.920000)(0.599113,0.930000)(0.644789,0.940000)(0.693639,0.950000)(0.746162,0.960000)
(0.802352,0.970000)(0.865232,0.980000)(1.000000,1.000000)

\newrgbcolor{color269.0145}{0  0  0}
\psline[plotstyle=line,linejoin=1,linestyle=dashed,linewidth=\LineWidth,linecolor=color269.0145]
(0.000000,0.529488)(0.010000,0.536265)(0.020000,0.543477)(0.030000,0.549352)(0.060000,0.569475)
(0.080000,0.582570)(0.100000,0.596434)(0.110000,0.603297)(0.120000,0.609706)(0.130000,0.615874)
(0.140000,0.622307)(0.150000,0.628366)(0.160000,0.636637)(0.170000,0.643090)(0.180000,0.648897)
(0.200000,0.662131)(0.210000,0.670139)(0.230000,0.681867)(0.240000,0.689278)(0.250000,0.694568)
(0.260000,0.702508)(0.270000,0.707983)(0.280000,0.714732)(0.290000,0.719876)(0.300000,0.727008)
(0.310000,0.733434)(0.320000,0.739162)(0.330000,0.745563)(0.340000,0.752399)(0.350000,0.758699)
(0.360000,0.763848)(0.380000,0.775237)(0.390000,0.781793)(0.400000,0.786412)(0.410000,0.792232)
(0.420000,0.798436)(0.440000,0.809823)(0.450000,0.815285)(0.460000,0.820501)(0.470000,0.827259)
(0.480000,0.833084)(0.500000,0.844031)(0.520000,0.853755)(0.530000,0.859705)(0.540000,0.864518)
(0.550000,0.869765)(0.560000,0.875380)(0.570000,0.879773)(0.580000,0.885077)(0.600000,0.894438)
(0.610000,0.899612)(0.620000,0.903345)(0.630000,0.908530)(0.640000,0.912100)(0.650000,0.917497)
(0.670000,0.925176)(0.680000,0.929375)(0.690000,0.933878)(0.700000,0.937301)(0.710000,0.941645)
(0.720000,0.943954)(0.730000,0.948451)(0.750000,0.955345)(0.760000,0.957937)(0.780000,0.964919)
(0.790000,0.967518)(0.800000,0.969897)(0.810000,0.972834)(0.830000,0.977970)(0.840000,0.980015)
(0.860000,0.984768)(0.870000,0.986244)(0.900000,0.991651)(0.930000,0.995747)(0.950000,0.997707)
(0.970000,0.999108)(0.980000,0.999619)(0.990000,0.999914)

{ \scriptsize  
\rput(0.5,0.35){%
\psshadowbox[framesep=0pt,shadowsize=0pt,linewidth=\AxesLineWidth]{\psframebox*{\begin{tabular}{l}
\Rnode{a1}{\hspace*{0.0ex}} \hspace*{0.7cm} \Rnode{a2}{~~Check node} \\
\Rnode{a3}{\hspace*{0.0ex}} \hspace*{0.7cm} \Rnode{a4}{~~Variable node} \\
\end{tabular}}
\ncline[linestyle=solid,linewidth=\LineWidth,linecolor=color268.015]{a1}{a2}
\ncline[linestyle=dashed,linewidth=\LineWidth,linecolor=color269.0145]{a3}{a4}
}%
}%
} 

{ \small 
\newrgbcolor{color200.0145}{0  0  0}
\uput{0pt}[0](0.4855,0.755418){%
\psframebox[framesep=1pt,fillstyle=solid,linestyle=none,linewidth=0.5pt]{\begin{tabular}{@{}c@{}}
$\sigma^2_d = 3$\\[-0.3ex]
\end{tabular}}}
} 

\end{pspicture}%
\end{minipage}
\caption{EXIT charts of $(3,6)$ regular LDPC code resulting from \textbf{Algorithm \ref{alg:exit}} for SNR$=3$ dB and different decoder noise variances $\sigma^2_d$.}
\label{fig:exits}
\end{figure*}
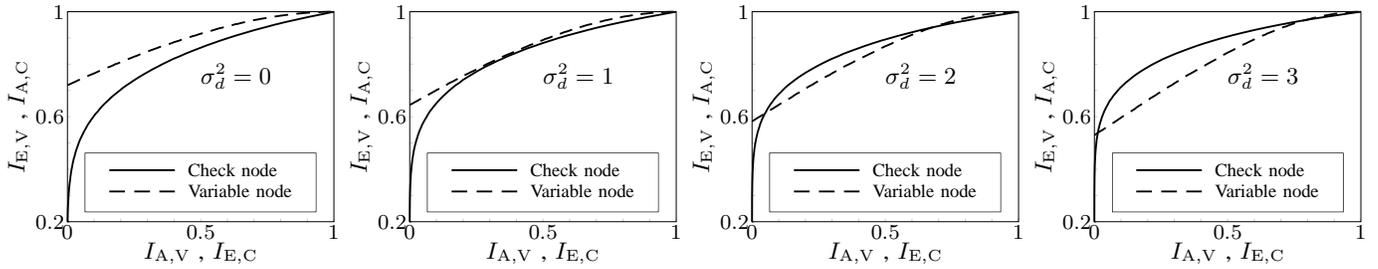
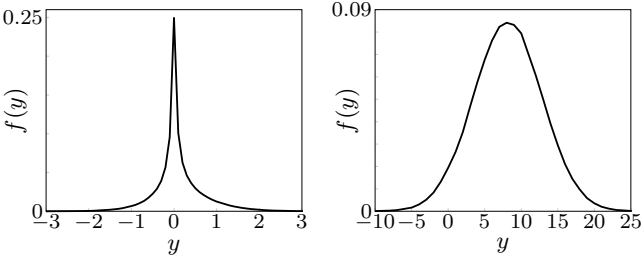
\begin{figure}[t]
\centering
\begin{minipage}[b]{0.48\columnwidth}
\setlength{\plotwidth}{0.8\columnwidth}
%
%
\psset{xunit=0.166667\plotwidth,yunit=3.033499\plotwidth}%
\begin{pspicture}(-3.760369,-0.028889)(3.069124,0.260000)%


\psline[linewidth=\AxesLineWidth,linecolor=GridColor](-3.000000,0.000000)(-3.000000,0.003956)
\psline[linewidth=\AxesLineWidth,linecolor=GridColor](-2.000000,0.000000)(-2.000000,0.003956)
\psline[linewidth=\AxesLineWidth,linecolor=GridColor](-1.000000,0.000000)(-1.000000,0.003956)
\psline[linewidth=\AxesLineWidth,linecolor=GridColor](0.000000,0.000000)(0.000000,0.003956)
\psline[linewidth=\AxesLineWidth,linecolor=GridColor](1.000000,0.000000)(1.000000,0.003956)
\psline[linewidth=\AxesLineWidth,linecolor=GridColor](2.000000,0.000000)(2.000000,0.003956)
\psline[linewidth=\AxesLineWidth,linecolor=GridColor](3.000000,0.000000)(3.000000,0.003956)
\psline[linewidth=\AxesLineWidth,linecolor=GridColor](-3.000000,0.000000)(-2.928000,0.000000)
\psline[linewidth=\AxesLineWidth,linecolor=GridColor](-3.000000,0.050000)(-2.928000,0.050000)
\psline[linewidth=\AxesLineWidth,linecolor=GridColor](-3.000000,0.100000)(-2.928000,0.100000)
\psline[linewidth=\AxesLineWidth,linecolor=GridColor](-3.000000,0.150000)(-2.928000,0.150000)
\psline[linewidth=\AxesLineWidth,linecolor=GridColor](-3.000000,0.200000)(-2.928000,0.200000)
\psline[linewidth=\AxesLineWidth,linecolor=GridColor](-3.000000,0.250000)(-2.928000,0.250000)

{ \footnotesize 
\rput[t](-3.000000,-0.003956){$-3$}
\rput[t](-2.000000,-0.003956){$-2$}
\rput[t](-1.000000,-0.003956){$-1$}
\rput[t](0.000000,-0.003956){$0$}
\rput[t](1.000000,-0.003956){$1$}
\rput[t](2.000000,-0.003956){$2$}
\rput[t](3.000000,-0.003956){$3$}
\rput[r](-3.072000,0.000000){$0$}
\rput[r](-3.072000,0.250000){$0.25$}
} 

\psframe[linewidth=\AxesLineWidth,dimen=middle](-3.000000,0.000000)(3.000000,0.260000)

{ \small 
\rput[b](0.000000,-.06){
\begin{tabular}{c}
$y$\\
\end{tabular}
}

\rput[t]{90}(-4,0.130000){
\begin{tabular}{c}
$f(y)$\\
\end{tabular}
}
} 

\newrgbcolor{color188.0028}{0  0  0}
\psline[plotstyle=line,linejoin=1,linestyle=solid,linewidth=\LineWidth,linecolor=color188.0028]
(-3.000000,0.000010)(-3.000000,0.000010)
\psline[plotstyle=line,linejoin=1,linestyle=solid,linewidth=\LineWidth,linecolor=color188.0028]
(3.000000,0.000213)(3.000000,0.000213)
\psline[plotstyle=line,linejoin=1,linestyle=solid,linewidth=\LineWidth,linecolor=color188.0028]
(-3.000000,0.000010)(-2.200000,0.000258)(-1.900000,0.000627)(-1.500000,0.001980)(-1.300000,0.003145)
(-1.200000,0.004068)(-1.000000,0.006697)(-0.900000,0.008290)(-0.800000,0.010642)(-0.700000,0.013377)
(-0.600000,0.017440)(-0.500000,0.022493)(-0.400000,0.028965)(-0.300000,0.039062)(-0.200000,0.056637)
(-0.100000,0.095897)(0.000000,0.249352)(0.100000,0.100905)(0.200000,0.062575)(0.300000,0.046103)
(0.400000,0.036598)(0.500000,0.029707)(0.600000,0.024655)(0.700000,0.020827)(0.800000,0.017585)
(0.900000,0.014780)(1.000000,0.012498)(1.300000,0.007535)(1.400000,0.006293)(1.600000,0.004298)
(1.800000,0.002883)(2.100000,0.001557)(2.300000,0.000952)(2.900000,0.000197)(3.000000,0.000213)

\end{pspicture}%

\end{minipage}
\begin{minipage}[b]{0.48\columnwidth}
\setlength{\plotwidth}{0.8\columnwidth}
%
%
\psset{xunit=0.028571\plotwidth,yunit=8.763441\plotwidth}%
\begin{pspicture}(-14.435484,-0.010000)(25.645161,0.092105)%


\psline[linewidth=\AxesLineWidth,linecolor=GridColor](-10.000000,0.000000)(-10.000000,0.001369)
\psline[linewidth=\AxesLineWidth,linecolor=GridColor](-5.000000,0.000000)(-5.000000,0.001369)
\psline[linewidth=\AxesLineWidth,linecolor=GridColor](0.000000,0.000000)(0.000000,0.001369)
\psline[linewidth=\AxesLineWidth,linecolor=GridColor](5.000000,0.000000)(5.000000,0.001369)
\psline[linewidth=\AxesLineWidth,linecolor=GridColor](10.000000,0.000000)(10.000000,0.001369)
\psline[linewidth=\AxesLineWidth,linecolor=GridColor](15.000000,0.000000)(15.000000,0.001369)
\psline[linewidth=\AxesLineWidth,linecolor=GridColor](20.000000,0.000000)(20.000000,0.001369)
\psline[linewidth=\AxesLineWidth,linecolor=GridColor](25.000000,0.000000)(25.000000,0.001369)
\psline[linewidth=\AxesLineWidth,linecolor=GridColor](-10.000000,0.000000)(-9.580000,0.000000)
\psline[linewidth=\AxesLineWidth,linecolor=GridColor](-10.000000,0.010000)(-9.580000,0.010000)
\psline[linewidth=\AxesLineWidth,linecolor=GridColor](-10.000000,0.020000)(-9.580000,0.020000)
\psline[linewidth=\AxesLineWidth,linecolor=GridColor](-10.000000,0.030000)(-9.580000,0.030000)
\psline[linewidth=\AxesLineWidth,linecolor=GridColor](-10.000000,0.040000)(-9.580000,0.040000)
\psline[linewidth=\AxesLineWidth,linecolor=GridColor](-10.000000,0.050000)(-9.580000,0.050000)
\psline[linewidth=\AxesLineWidth,linecolor=GridColor](-10.000000,0.060000)(-9.580000,0.060000)
\psline[linewidth=\AxesLineWidth,linecolor=GridColor](-10.000000,0.070000)(-9.580000,0.070000)
\psline[linewidth=\AxesLineWidth,linecolor=GridColor](-10.000000,0.080000)(-9.580000,0.080000)
\psline[linewidth=\AxesLineWidth,linecolor=GridColor](-10.000000,0.090000)(-9.580000,0.090000)

{ \footnotesize 
\rput[t](-10.000000,-0.001369){$-10$}
\rput[t](-5.000000,-0.001369){$-5$}
\rput[t](0.000000,-0.001369){$0$}
\rput[t](5.000000,-0.001369){$5$}
\rput[t](10.000000,-0.001369){$10$}
\rput[t](15.000000,-0.001369){$15$}
\rput[t](20.000000,-0.001369){$20$}
\rput[t](25.000000,-0.001369){$25$}
\rput[r](-10.420000,0.000000){$0$}
\rput[r](-10.420000,0.090000){$0.09$}
} 

\psframe[linewidth=\AxesLineWidth,dimen=middle](-10.000000,0.000000)(25.000000,0.090000)

{ \small 
\rput[b](7.500000,-0.02){
\begin{tabular}{c}
$y$\\
\end{tabular}
}

\rput[t]{90}(-15.5,0.045000){
\begin{tabular}{c}
$f(y)$\\
\end{tabular}
}
} 

\newrgbcolor{color188.0034}{0  0  0}
\psline[plotstyle=line,linejoin=1,linestyle=solid,linewidth=\LineWidth,linecolor=color188.0034]
(25.000000,0.000153)(25.000000,0.000153)
\psline[plotstyle=line,linejoin=1,linestyle=solid,linewidth=\LineWidth,linecolor=color188.0034]
(-10.000000,0.000087)(-8.000000,0.000297)(-7.000000,0.000520)(-5.000000,0.001630)(-4.000000,0.003023)
(-3.000000,0.005160)(-2.000000,0.008363)(-1.000000,0.013250)(0.000000,0.019430)(1.000000,0.026377)
(2.000000,0.035273)(3.000000,0.046727)(4.000000,0.057687)(5.000000,0.067643)(6.000000,0.076183)
(7.000000,0.082340)(8.000000,0.084170)(9.000000,0.083187)(10.000000,0.079443)(12.000000,0.060537)
(13.000000,0.049990)(14.000000,0.039133)(15.000000,0.029503)(16.000000,0.020987)(17.000000,0.014450)
(18.000000,0.009737)(19.000000,0.005870)(20.000000,0.003560)(21.000000,0.002017)(22.000000,0.001167)
(23.000000,0.000577)(25.000000,0.000153)

\end{pspicture}%

\end{minipage}
\caption{\revise{Empirical distributions of outputs of check (left) and variable (right) nodes for $I_A=0.5$, $\sigma_d^2=1$ and SNR $=3$ dB.}}
\label{fig:dist}
\end{figure}

To obtain the EXIT curves for the noisy LDPC decoder, we use \textbf{Algorithm \ref{alg:exit}} and the empirical distributions computed by running simulations for each degree of variable node and check node.
Specifically for each decoder component, we compute the extrinsic mutual information ${I}_\mathrm{E}$ corresponding to a priori mutual information ${I}_\mathrm{A}$ for the noisy variable (check) node decoder. To maintain the desired structure of the iterative decoder \cite{ten04} for the noisy decoder, we modified the variable nodes and check nodes as shown in Fig. \ref{modnode}.
In fact, we have considered decoder noise as a part of variable (check) nodes and call them noisy variable (check) node decoders, NVND (NCND).

In \textbf{Algorithm \ref{alg:exit}}, the proposed method for finding the EXIT curve of NVND is described. The procedure of finding EXIT curves for the noisy check node is similar to that of the noisy variable node; however, for the check node the communication channel output is not fed to the check node, i.e., the only input of the noisy check node is \revise{the vector of a priori LLRs from variable nodes (${\bf AP}$}). The presented algorithm is an extension of the algorithm 7.4 in \cite{sara} for Turbo codes.

\revise{It is noteworthy that for a noiseless decoder, input messages to each decoder component (VND or CND) are modeled by Gaussian random variables of mean $\mu_A$ and variance $\sigma^2_A=2\mu_A$. Then the mutual information $I_A$ of each input message and the channel input $X$ (a priori mutual information) can be computed using \[I_A=J\left(\sigma_A\right)\,.\] This relation implies a one-to-one mapping from each $I_A$ to a $\sigma_A$ (and consequently $\mu_A$). However, for a noisy decoder there is not such a one-to-one mapping and for each $I_A$ different sets of $(\mu_A,\sigma^2_A)$ may be found. One way to tackle this problem is to assume consistency for input messages to the noisy components (NVND and NCND) as stated in step 6 of \textbf{Algorithm \ref{alg:exit}}, and calculate each a priori LLR as in step 7. Then, the input messages to VND and CND are no longer consistent (since they are corrupted with decoder noise). Since we simulate each decoder component separately, the assumption made does not affect the computation of EXIT curves for the other components, and as our experiments validate, the suggested approach provides very accurate EXIT curves.
}

Fig.~\ref{fig:exits} illustrates the evolution of EXIT charts of $(3,6)$ regular LDPC code for different values of decoder noise variance $\sigma^2_d$ and for communication channel SNR $3$ dB.
For $\sigma^2_d=0$, there is an open tunnel between the curves, and increasing $\sigma^2_d$ to one makes the tunnel tighter.
However, for $\sigma^2_d =2$ and $\sigma^2_d=3$ the variable and the check EXIT curves cross each other.
The EXIT charts in Fig.~\ref{fig:exits} illustrate that the SNR threshold of $(3,6)$ regular LDPC code is less than $3$ dB when $\sigma^2_d=0$ and $\sigma^2_d=1$ and is greater than $3$ dB when $\sigma^2_d=2$ and $\sigma^2_d=3$.
This is in line with the results obtained from the density evolution analysis, as shown in Table \ref{table1}.

Fig.~\ref{fig:dist} depicts the empirical distributions $f(y)$ for the outputs of NVND of degree $d_v=3$ and NCND of degree $d_c=6$, used in \eqref{MI} to find the extrinsic mutual information $I_\mathrm{E}$. We observe that the distribution of the nodes' outputs are light-tailed for the check node and normal-tailed for the variable node. This motivates the use of numerical integration in \eqref{MI} by integrating over a limited integral domain.

In the following \revise{section}, we will use the EXIT curves resulting from \textbf{Algorithm \ref{alg:exit}} to design robust LDPC codes for noisy decoders.
\section{Design of Robust Irregular LDPC Codes for Noisy Decoder}\label{sec:design}
In this \revise{section}, our goal is to design robust irregular LDPC codes for noisy decoder.
The design procedure of LDPC codes \revise{is based on fitting EXIT curves corresponding to given variable and check degree distributions.}
In \cite{ten04}, for a noiseless decoder, right-regular LDPC codes are designed for a fixed check node degree of $d_{c}$, \revise{by} fitting a weighted sum of the EXIT curves of variable nodes to the check node EXIT curve.
In this work, we design irregular LDPC codes and allow more than one degree for both variable \revise{nodes} and check nodes.
Our benchmark for comparisons are irregular LDPC codes with the same rates designed for a noiseless decoder in \cite{rich01}.
We verify the performance of the designed codes by \revise{the} simulations of finite-length LDPC codes.

It is noteworthy that when the exchanged messages in the decoder are not consistent, the weighted sum of the EXIT chart curves of variable node and check node are not the exact curves of the irregular code. \revise{However}, our simulation results indicate that this approach is accurate enough for the case with a noisy decoder.

\subsection{Code Design Algorithm}
Using the EXIT curves obtained from {\bf Algorithm \ref{alg:exit}},
we propose a simple method for the design of irregular codes for the noisy decoder.
Similar to the code design approach proposed in \cite{rich01},
we restrict our attention to irregular codes with two consecutive check node degrees as \begin{equation}\rho(x)={\alpha}x^{d_{c}-1}+(1-{\alpha})x^{d_{c}},\label{eq:cdegree}\end{equation}
and variable node degrees as \begin{equation}\lambda(x)=\sum_{i=2}^{D_{v}}\lambda_{i}x^{i-1}.\label{eq:vdegree}\end{equation}
The \textit{effective} EXIT curve of a noisy irregular check (variable) node $I_\mathrm{A,NCND}$ $(I_\mathrm{E,NVND})$ is obtained by averaging over the EXIT curves of check (variable) nodes with constituent check (node) degrees \cite{ashikh04}.

In this paper, we refer to a candidate code $\mathcal{C}=\left(\rho(x),\lambda(x)\right)$ of rate $r$ as a \textit{successful code} for a given SNR, if
it satisfies the following constraints
\begin{equation}\begin{split}
i)&\quad \lambda(1)=1\,,\\
ii)&\quad \rho(1)=1\,,\\
iii)&\quad r=1-\frac{\int_{0}^{1}\rho(x)dx}{\int_{0}^{1}\lambda(x)dx}\,,\\
iv)&\quad I_\mathrm{E,NVND} \succ I_\mathrm{A,NCND}\,,
\label{eq:crit}\end{split}\end{equation}
where the last constraint implies that the effective EXIT curve of its variable node lies above the effective EXIT curve of its check node.

For a given code rate $r$, we are looking for the code ${\mathcal{C}^*=\left(\rho(x)^*,\lambda(x)^*\right)}$ corresponding to the threshold SNR, $\textnormal{SNR}_\mathrm{th}$, i.e., the minimum SNR for which there is a successful code.
\begin{algorithm}[t]
\begin{algorithmic}[1]
\caption{Robust LDPC code design}
\label{alg:design}
\State \textbf{Input:} $d_c$, $D_v$, $r$, $\sigma^2_d$, $\Delta$, ${\mathbf{S}}$
\State \textbf{Output:} code $\mathcal{C}^*=\left(\rho(x)^*,\lambda(x)^*\right)$
\State {initialize $\text{SNR}_\mathrm{th}$}
\State { $\sigma^2_n=\left(2r\text{SNR}_\mathrm{th}\right)^{-1}$}
\For {$i = 1:M+1$}
\State {$\alpha= \mathbf{S}(i)$}\\
\quad \quad{[compute EXIT curves of check nodes with degree \nonumber\\  \quad \quad ${d_c-1}$ and degree $d_c$ using \textbf{Algorithm \ref{alg:exit}}]}\\

\quad \quad{[compute effective EXIT curve $I_{\mathrm{A,NCND}}$ using $\rho(x)$ \\ \quad \quad in \eqref{eq:cdegree}]}\\

\quad \quad {[compute EXIT curves of variable nodes with degrees \\ \quad \quad $d_v=2$ to $d_v=D_v$ using \textbf{Algorithm \ref{alg:exit}}]}\\

\quad \quad {[check if there is a $\lambda(x)$ such that $\mathcal{C}=\left(\rho(x),\lambda(x)\right)$ \\ \quad \quad satisfies the constraints in (25)]}
\If {code $\mathcal{C}$ exists}
\State {$\text{SNR}_{\mathrm{th}} = \text{SNR}_{\mathrm{th}}-\Delta$}
\State {$\mathcal{C}^*=\mathcal{C}$}
\State {go to step 4}
\EndIf
\EndFor
\end{algorithmic}
\end{algorithm}
In \textbf{Algorithm \ref{alg:design}}, we set the check node degree $d_c$, the maximum variable node degree $D_v$, and design rate $r$.
\revise{In order to speed up the design procedure}, we let $\alpha$ take limited values in $\mathbf{S}=[0:1/M:1]$.
Then, for a given $\text{SNR}_\mathrm{th}$, by varying $\alpha$ from zero to one, we check if there is a variable degree distribution $\lambda(x)$, as in \eqref{eq:vdegree}, such that the constraints in \eqref{eq:crit} are satisfied. \revise{One way to do this is for each $\alpha$ to find $I_\mathrm{A,NCND}$ according to \eqref{eq:cdegree} and check if there is a vector $(\lambda_2,\ldots,\lambda_{D_v})$ for which the constraints in \eqref{eq:crit} are feasible. 
If so, we form the code $\mathcal{C}=\left(\rho(x),\lambda(x)\right)$. Subsequently, we reduce $\text{SNR}_\mathrm{th}$ by $\Delta$ and the algorithm does another iteration; otherwise, it terminates and the successful code from the previous iteration is selected.}

The EXIT curves of a check node do not depend on $\text{SNR}_\mathrm{th}$; By changing $\text{SNR}_\mathrm{th}$, the EXIT curves of variable nodes, and consequently $I_{\mathrm{E,NVND}}$ are affected.
Using this fact, the complexity of the proposed algorithm may be further reduced by pre-computing $I_\mathrm{A,NCND}$ for each value of $\alpha$ once and storing them before running the algorithm.
In the following \revise{section}, we will present some results illustrating the application of the proposed algorithm in the design of robust LDPC codes for noisy decoders.

\subsection{Design Examples}
From Table 1 of \cite{rich01}, for the maximum variable degree of four, the introduced code of rate one-half has two types of check nodes with degrees five and six. The threshold of this code is $0.8085$ dB and has \revise{the following degree distributions}
\begin{align*}
&\lambda(x)=0.384x+0.042x^{2}+0.574x^{3},\\
&\rho(x)=0.241x^{4}+0.759x^{5}.
\end{align*}
However, when the decoder is not perfect, the threshold and also the performance of this code degrade. For instance, in noisy decoder the threshold has increased by about $1.7$ dB and $2.5$ dB for decoder noise variances $\sigma^{2}_{d}=0.5$ and $\sigma^{2}_{d}=1$, respectively.

Considering the same constraints in check degree distribution and maximum variable degree, we have designed an irregular one-half rate LDPC code for the noisy decoder with $\sigma^{2}_{d}=0.5$.
Using \textbf{Algorithm \ref{alg:design}}, we obtained \textit{code A} with the following degree distributions
\begin{align*}
&\lambda(x)=0.453x+0.547x^{3},\\
&\rho(x)=0.451x^{4}+0.549x^{5}\,.
\end{align*}

\begin{figure}[t]
\input{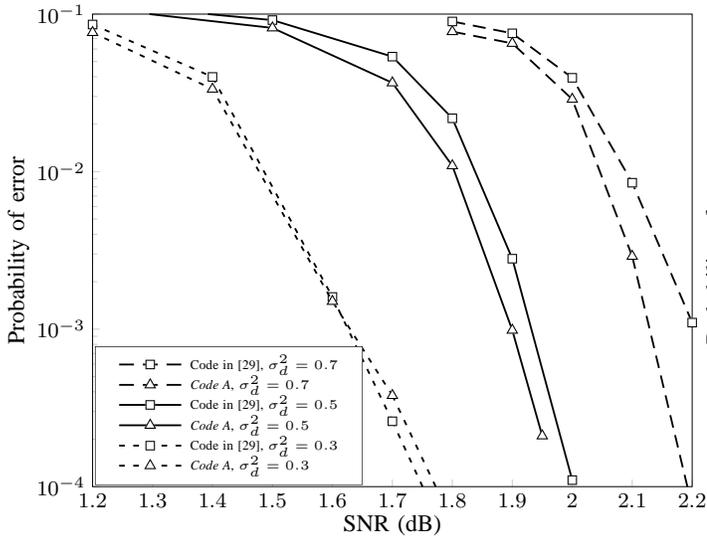}
\centering
\caption{\revise{Error probability performances of \textit{Code A} and the code in [29] for different decoder noise variances, $N=10^4$.}}
\label{design_half}
\end{figure}
Fig.~\ref{design_half} shows the performance of \textit{Code A} and the designed code in \cite{rich01} for different decoder noise variances. When $\sigma^2_d=0.5$ (solid curves), \textit{Code A} provides a better performance compared to the code in \cite{rich01}. By increasing the decoder noise variance to $\sigma^2_d=0.7$ (dashed curves) \textit{Code A} still outperforms the code in \cite{rich01}, while by decreasing the decoder noise variance to $\sigma^2_d=0.3$ (dotted curves) the codes show almost the same performance. It is evident that \textit{Code A} performs well when the decoder noise power varies in the proximity of the designed variance.
It is noteworthy that in our simulations the Tanner graph of codes are free from cycles of length four. The maximum number of decoder iterations is set to $80$, and for each SNR, the error probability is computed after a total number of $50$ block errors occur.

As another example of code design, considering the same constraints on degree distributions, we have designed an irregular code (\textit{Code B}) for noisy decoder with $\sigma^{2}_{d}=1$. The degree \revise{distributions} of this code \revise{are}
\begin{align*}
&\lambda(x)=0.4808x+0.5192x^{3},\\
&\rho(x)=0.553x^{4}+0.447x^{5}.
\end{align*}
Fig.~\ref{design_1} shows the simulation results of \textit{Code B} and the designed code in [29] for different decoder noise variances. When $\sigma^2_d=1$ (solid curves), \textit{Code B} shows better performance compared to the code in [29]. Similar observations are made when the decoder noise variance (dashed curves) increases to $1.2$ or decreases to $0.8$ (dotted curves).

%

In the above designs, we considered two consecutive check degrees for a fair comparison with the code in \cite{rich01}. However, the presented approach may be used to design irregular LDPC codes with arbitrary check degrees as \[\rho=\sum_{i=2}^{D_c}\rho_ix^{i-1}\,.\]
The constraints in \eqref{eq:crit} remain linear and hence the same procedure still applies. We emphasize that the range of decoder internal noise power that we considered here (between $0$ and $1.2$) is rather conservative. For a scalar quantization of a Gaussian random variable, a single bit change in the quantization bitrate scales the quantization noise variance by a factor of about $3.4$ \cite{Sayood}. According to our simulation results, the proposed robust LDPC code design indicates a higher gain when the decoder internal noise power is stronger.

\section{Conclusions}\label{sec:conclude}
We considered a noisy \revise{message-passing} scheme for the decoding of LDPC codes over AWGN communication channels. We modeled the internal decoder noise as additive white Gaussian and observed the inconsistency of the exchanged message densities in the iterative decoder.
For the inconsistent LDPC decoder, a density evolution scheme was formulated to track both the mean and the variance of the messages exchanged in the decoder.
We quantified the increase of the decoding threshold SNR as a consequence of the internal decoder noise based on a density evolution analysis. We also analyzed the performance of the noisy decoder using EXIT charts. We introduced an algorithm based on the computed EXIT charts to design robust irregular LDPC codes. The designed codes partially compensate the performance loss due to the decoder internal noise.
In this work, we modeled the decoder noise as AWGN on the exchanged messages, however, an interesting future step is to incorporate other noise models possibly directly obtained from practical decoder implementations. One may also use this research to design codes that are robust to decoder internal noise whose power may vary in a given range depending on possible types of implementation.

\begin{figure}[t]
\input{RichB.tex}
\centering
\caption{\revise{Error probability performances of \textit{Code B} and the code in \cite{rich01} for different decoder noise variances, $N=10^4$.}}
\label{design_1}
\end{figure}

\ifCLASSOPTIONcaptionsoff
  \newpage
\fi

\bibliographystyle{IEEEtran}
\bibliography{IEEEabrv,ref}

\begin{thebibliography}{10}
\providecommand{\url}[1]{#1}
\csname url@samestyle\endcsname
\providecommand{\newblock}{\relax}
\providecommand{\bibinfo}[2]{#2}
\providecommand{\BIBentrySTDinterwordspacing}{\spaceskip=0pt\relax}
\providecommand{\BIBentryALTinterwordstretchfactor}{4}
\providecommand{\BIBentryALTinterwordspacing}{\spaceskip=\fontdimen2\font plus
\BIBentryALTinterwordstretchfactor\fontdimen3\font minus
  \fontdimen4\font\relax}
\providecommand{\BIBforeignlanguage}[2]{{%
\expandafter\ifx\csname l@#1\endcsname\relax
\typeout{** WARNING: IEEEtran.bst: No hyphenation pattern has been}%
\typeout{** loaded for the language `#1'. Using the pattern for}%
\typeout{** the default language instead.}%
\else
\language=\csname l@#1\endcsname
\fi
#2}}
\providecommand{\BIBdecl}{\relax}
\BIBdecl

\bibitem{Gallag2}
R.~Gallager, ``{L}ow-density parity-check codes,'' \emph{IRE Trans. Inf.
  Theory}, vol.~8, no.~1, pp. 21--28, Jan 1962.

\bibitem{spil}
M.~Sipser and D.~Spielman, ``Expander codes,'' \emph{IEEE Trans. Inf. Theory},
  vol.~42, no.~6, pp. 1710--1722, Nov 1996.

\bibitem{mackay1}
D.~J. MacKay and R.~M. Neal, ``Near shannon limit performance of low density
  parity check codes,'' \emph{Electronics letters}, vol.~32, no.~18, pp.
  1645--1646, 1996.

\bibitem{mackay2}
D.~J. MacKay, ``Good error-correcting codes based on very sparse matrices,''
  \emph{IEEE Trans. Inf. Theory}, vol.~45, no.~2, pp. 399--431, 1999.

\bibitem{dvb}
M.~Eroz, F.~Sun, and L.~Lee, ``{DVB-S2} low-density parity-check codes with
  near shannon limit performance,'' \emph{Int. Journal of Satellite
  Communications and Networking}, vol.~22, no.~3, 2004.

\bibitem{std}
``Part 16: {A}ir interface for fixed and mobile broadband wireless access
  systems amendment for physical and medium access control layers for combined
  fixed and mobile operation in licensed bands,'' \emph{IEEE P802.16e}, Oct
  2004.

\bibitem{chung01}
S.-Y. Chung, T.~J. Richardson, and R.~L. Urbanke, ``Analysis of sum-product
  decoding of low-density parity-check codes using a {G}aussian
  approximation,'' \emph{IEEE Trans. Inf. Theory}, vol.~47, no.~2, pp.
  657--670, 2001.

\bibitem{ten04}
S.~ten Brink, G.~Kramer, and A.~Ashikhmin, ``Design of low-density parity-check
  codes for modulation and detection,'' \emph{IEEE Trans. Commun.}, vol.~52,
  no.~4, pp. 670--678, April 2004.

\bibitem{ashikh04}
A.~Ashikhmin, G.~Kramer, and S.~ten Brink, ``Extrinsic information transfer
  functions: model and erasure channel properties,'' \emph{IEEE Trans. Inf.
  Theory}, vol.~50, no.~11, pp. 2657--2673, Nov 2004.

\bibitem{rich001}
T.~J. Richardson and R.~L. Urbanke, ``The capacity of low-density parity-check
  codes under message-passing decoding,'' \emph{IEEE Trans. Inf. Theory},
  vol.~47, no.~2, pp. 599--618, Feb 2001.

\bibitem{loeli01}
H.-A. Loeliger, F.~Lustenberger, M.~Helfenstein, and F.~Tark{\"o}y,
  ``Probability propagation and decoding in analog {VLSI},'' \emph{IEEE Trans.
  Inf. Theory}, vol.~47, no.~2, pp. 837--843, 2001.

\bibitem{pego06}
T.~Pegoraro, F.~Gomes, R.~Lopes, R.~Gallo, J.~Panaro, M.~Paiva, F.~Oliveira,
  and F.~Lumertz, ``Design, simulation and hardware implementation of a digital
  television system: {LDPC} channel coding,'' in \emph{IEEE 9th Int. Symp.
  Spread Spectrum Techniques and Applications}, Aug 2006, pp. 203--207.

\bibitem{zhang07}
Z.~Zhang, L.~Dolecek, M.~Wainwright, V.~Anantharam, and B.~Nikolic,
  ``Quantization effects in low-density parity-check decoders,'' in \emph{IEEE
  Int. Conf. Commun. (ICC)}, June 2007, pp. 6231--6237.

\bibitem{lee96}
D.~Lee and D.~Neuhoff, ``Asymptotic distribution of the errors in scalar and
  vector quantizers,'' \emph{IEEE Trans. Inf. Theory}, vol.~42, no.~2, pp.
  446--460, Mar 1996.

\bibitem{varsh11}
L.~Varshney, ``Performance of {LDPC} codes under faulty iterative decoding,''
  \emph{IEEE Trans. Inf. Theory}, vol.~57, no.~7, pp. 4427--4444, July 2011.

\bibitem{leduc15}
F.~Leduc-Primeau, F.~R. Kschischang, and W.~J.~Gross, ``Modeling and energy
  optimization of {LDPC} decoder circuits with timing violations,'' available
  online: arxiv.org/pdf/1503.03880.pdf.

\bibitem{hagen2}
J.~Hagenauer and M.~Winklhofer, ``The analog decoder,'' in \emph{IEEE Int.
  Symp. Inf. Theory}, Aug 1998, p. 145.

\bibitem{koch09}
T.~Koch, A.~Lapidoth, and P.~Sotiriadis, ``Channels that heat up,'' \emph{IEEE
  Trans. Inf. Theory}, vol.~55, no.~8, pp. 3594--3612, Aug 2009.

\bibitem{Hsi15}
C.-H. Huang, Y.~Li, and L.~Dolecek, ``Belief propagation algorithms on noisy
  hardware,'' \emph{IEEE Trans. Commun.}, vol.~63, no.~1, pp. 11--24, Jan 2015.

\bibitem{Taba12}
S.~Tabatabaei~Yazdi, C.-H. Huang, and L.~Dolecek, ``Optimal design of a
  {G}allager {B} noisy decoder for irregular {LDPC} codes,'' \emph{IEEE Commun.
  Lett.}, vol.~16, no.~12, pp. 2052--2055, Dec 2012.

\bibitem{Taba13}
S.~Tabatabaei~Yazdi, H.~Cho, and L.~Dolecek, ``{G}allager {B} decoder on noisy
  hardware,'' \emph{IEEE Trans. Commun.}, vol.~61, no.~5, pp. 1660--1673, May
  2013.

\bibitem{Huang14}
C.-H. Huang, Y.~Li, and L.~Dolecek, ``{G}allager {B} {LDPC} decoder with
  transient and permanent errors,'' \emph{IEEE Trans. Commun.}, vol.~62, no.~1,
  pp. 15--28, Jan 2014.

\bibitem{us10}
A.~Tarighati, H.~Farhadi, and F.~Lahouti, ``Performance analysis of noisy
  message-passing decoding of low-density parity-check codes,'' in \emph{Proc.
  6th IEEE Int. Symp. Turbo Codes and Iterative Inform. Processing}, 2010.

\bibitem{moon}
T.~K. Moon, \emph{{E}rror {C}orrection {C}oding: {M}athematical {M}ethods and
  {A}lgorithms}.\hskip 1em plus 0.5em minus 0.4em\relax Wiley-Interscience,
  2005.

\bibitem{gamal01}
H.~El~Gamal and A.~R. Hammons~Jr, ``Analyzing the turbo decoder using the
  {G}aussian approximation,'' \emph{IEEE Trans. Inf. Theory}, vol.~47, no.~2,
  pp. 671--686, Feb 2001.

\bibitem{saeedi07}
H.~Saeedi and A.~H. Banihashemi, ``Performance of belief propagation for
  decoding {LDPC} codes in the presence of channel estimation error,''
  \emph{IEEE Trans. Commun.}, vol.~55, no.~1, pp. 83--89, Jan 2007.

\bibitem{ardak04}
M.~Ardakani and F.~R. Kschischang, ``A more accurate one-dimensional analysis
  and design of irregular {LDPC} codes,'' \emph{IEEE Trans. Commun.}, vol.~52,
  no.~12, pp. 2106--2114, Dec 2004.

\bibitem{ten01}
S.~ten Brink, ``Convergence behavior of iteratively decoded parallel
  concatenated codes,'' \emph{IEEE Trans. Commun.}, vol.~49, no.~10, pp.
  1727--1737, Oct 2001.

\bibitem{rich01}
T.~J. Richardson, M.~A. Shokrollahi, and R.~L. Urbanke, ``Design of
  capacity-approaching irregular low-density parity-check codes,'' \emph{IEEE
  Trans. Inf. Theory}, vol.~47, no.~2, pp. 619--637, Feb 2001.

\bibitem{hagen}
J.~Hagenauer, ``The {EXIT} chart--{I}ntroduction to extrinsic information
  transfer in iterative processing,'' in \emph{Eur. Signal Process. Conf.}, Sep
  2004, pp. 1541--1548.

\bibitem{sara}
S.~J. Johnson, \emph{Iterative {E}rror {C}orrection: {T}urbo, {L}ow-{D}ensity
  {P}arity-{C}heck {C}odes and {R}epeat-{A}ccumulate {C}odes}.\hskip 1em plus
  0.5em minus 0.4em\relax Cambridge University Press, Jan 2010.

\bibitem{Sayood}
K.~Sayood, \emph{Introduction to data compression}.\hskip 1em plus 0.5em minus
  0.4em\relax Newnes, 2012.

\end{thebibliography}

\end{document}